\documentclass[12pt]{article} 
\usepackage{jheppub}
\pdfoutput=1

\usepackage{graphicx}
\usepackage[numbers,sort&compress]{natbib}
\usepackage{amsmath}
\usepackage{amssymb}
\usepackage{marginnote}
\usepackage{relsize}
 
\title{Signals of a new phase in ${\mathcal N}=2$ gauge theory with a
  magnetic field on the three-sphere}

\author{Suphakorn Chunlen, Kasper Peeters, Pichet Vanichchapongjaroen\\
  and Marija Zamaklar}

\affiliation{Department of Mathematical Sciences,\\
Durham University,\\
South Road,\\
Durham DH1 3LE,\\
United Kingdom.}

\emailAdd{suphakorn.chunlen@durham.ac.uk} 
\emailAdd{kasper.peeters@durham.ac.uk} 
\emailAdd{pichet.vanichchapongjaroen@durham.ac.uk}
\emailAdd{marija.zamaklar@durham.ac.uk}

\abstract{We study the effect of a magnetic field on ${\mathcal N}=2$
  super-Yang-Mills on $S^3$ at strong coupling using the gauge/gravity
  correspondence. As in previous work that dealt with the theory in
  infinite volume, we find that increasing the magnetic field pushes
  the system into the confined phase. However, we in addition also
  find that, within the class of configurations with the same symmetry
  as those which describe the ground state at vanishing magnetic
  field, a mass gap appears in the spectrum.  This suggests the
  existence of a new phase with so far unexplored symmetry
  structure. We provide suggestions for the physical properties of
  this phase.}

\preprint{DCPT-14/15}
 
\begin{document}
\maketitle

\section{Introduction}
 
While a lot of effort has been invested in understanding the
properties of gauge theories in infinite volumes using holography,
much less attention has been spent so far on understanding holographic
duals in compact spaces.  In this paper we look at the effect of a
magnetic field on gauge theory in compact space in the simplest
setting in which the dual is $\text{AdS}_5 \times S^5$ supplemented
with probe D7-branes.

String theory in global $\text{AdS}_5\times S^5$ space is believed to
be dual to ${\mathcal N}=4$ super-Yang-Mills theory at zero
temperature, which lives on the boundary of this space, an $S^3\times
\mathbb{R}$. Adding probe D7-branes in this geometry corresponds to
adding flavour hypermultiplets to the system. In the absence of a
chemical potential or magnetic field, the asymptotic position of the
probe branes is an arbitrary parameter related to the bare quark
mass. It was first observed in \cite{Karch:2006bv} and later
investigated in detail in \cite{Karch:2009ph} that at zero
temperature, the probe brane undergoes a third order, topology
changing phase transition at some value of the bare quark mass.  There
are thus two phases, corresponding to what \cite{Filev:2012ch} calls
\emph{ball} branes and \emph{Minkowski} branes. Ball branes are those
branes that extend all the way to the origin of the AdS space, where
an $S^3 \in \text{AdS}_5$ wrapped by the probe brane shrinks to zero
size. When the bare quark mass is increased above some critical value,
Minkowski branes appear. The Minkowski branes do not reach the AdS
origin. At a finite distance from the AdS origin, the $S^3 \in S^5$
wrapped by the probe brane shrinks to zero size. See
figure~\ref{f:embeddings} below for a schematic summary. These two types of
branes obviously have different topologies, and their existence is
specific for the \emph{global} $\text{AdS}_5 \times S^5 $ space, as in
the infinite volume or Poincar\'e limit, only ball branes exist.

The physical reason for the existence of this phase transition at zero
temperature is the non-vanishing Casimir energy of the field theory on
the compact space. The compact sphere on which the theory lives
introduces a new scale in the problem, which is the size of the
sphere.  Therefore, all quantities, including the bare quark mass, are
now expressed in term of this scale.  One way of thinking about the
phase transition from Minkowski branes to ball
branes~\cite{Karch:2009ph} is that instead of decreasing the bare
quark mass, one keeps the quark mass of the probe fixed while
squeezing the system into an ever smaller volume. When the radius of
the sphere reaches the critical value $R_3^* = 0.18
\sqrt{\lambda}/m_q$ \cite{Karch:2009ph}, the system undergoes a phase
transition, in which ``mesonic'' particles ``fall apart'', since their
zero point energy due their confinement to the finite volume becomes
larger than their binding energy.

In a previous paper \cite{Chunlen:2012zy} we have looked at the
effects of an isospin chemical potential in this compact geometry,
using holographic methods. We have found that this system exhibits an
instability for sufficiently large values of the chemical
potential. In contrast to other related models, the first excitation
to condense is not a vector meson but rather a scalar charged under
the global $SO(4)$ symmetry group.  We have also explicitly
constructed the new ground state. There is thus reason to suspect that
other interesting things could happen due to the finite volume of the
system.
\medskip

In the present paper we would like to understand how the observed
phase transition between Minkowski and ball branes changes once a
magnetic field is introduced in the system.  This situation has
previously been analysed by~\cite{Filev:2012ch} but as we will show
explicitly, the magnetic field used in that paper does not satisfy all
equations of motion. Because of this, the results
of~\cite{Filev:2012ch} differ significantly from our findings.

We know that a magnetic field in general favours the formation of
bound states, so in that sense it has an effect which is opposite from
that of the compactness of the
space~\cite{Karch:2009ph,Filev:2012ch}. The question is thus how
these two competing effects modify the behaviour of the system. We
have found several interesting effects caused by the magnetic
field. First, the presence of a magnetic field on the brane worldvolume
leads to branes being ``repelled'' from the origin of the AdS space. A
similar effect was observed previously in the infinite volume,
Poincar\'e limit in~\cite{Erdmenger:2007bn}. Because the magnetic
field enhances the curving of branes away from the origin, it works in
favour of ball branes becoming Minkowski. This is indeed what we
observe, as for larger values of magnetic field fewer and fewer ball
branes remain present in the spectrum. To be precise, they are present
in the spectrum for some range of bare quark mass $0 <m_q
<M_{\text{ball}}(B)$, and the function $M_{\text{ball}}(B)$ is
monotonically decreasing as a function of the magnetic field. When the
magnetic field reaches some critical value $\hat{M}_1$, all ball
branes have disappeared from the spectrum.

A second, somewhat surprising feature caused by the magnetic field is
the appearance of a ``gap'' in the allowed values of the bare quark
mass.  Even for an infinitesimal value of the magnetic field we
observe that ball and Minkowski branes are separated by a range of
bare quark masses $\Delta(B)$ for which none of the two phases are
possible. This kind of behaviour was not observed before in the
infinite volume Poincar\'e limit. While we are confident that there
are no ball nor Minkowski branes present in the gap, this leaves open
the possibility that there are other solutions, with less symmetry,
that lead to a brane embedding with mass in the gap. In particular, we
will in section~\ref{s:fluctuations_and_gap} present some preliminary
results in the direction of constructing an inhomogeneous phase, as a
candidate for the ``missing'' gap branes.

\section{Review of global $\text{AdS}_5\times S^5$ and D7-brane phases}

In order to set the scene, we will in this section briefly review some
properties of the global $\text{AdS}_5\times S^5$ space.  We will also
review the various embeddings of D7-probe branes in these geometries in
the absence of external fields, as previously analysed
in detail in~\cite{Karch:2006bv,Karch:2009ph}.  The metric of the global
$\text{AdS}_5$ spaces at zero temperature is given by
\begin{equation}
\label{u-coor}
{\rm d}s^2 =-\frac{u^2}{R^2}\left(1+\frac{R^2}{4u^2}\right)^2\, {\rm d}t^2
       +u^2\left(1-\frac{R^2}{4u^2}\right)^2 {\rm d}\bar{\Omega}_3^2
       +\frac{R^2}{u^2}\left( {\rm d}u^2+ u^2 {\rm d} \Omega_5^2 \right) \, , 
\end{equation}
where
\begin{equation}
{\rm d} \Omega_5^2 = \frac{{\rm d}\chi^2}{1-\chi^2}+\chi^2 {\rm
  d}\kappa^2+ (1-\chi^2) {\rm d}\Omega_3^2
\end{equation}
and 
\begin{equation}
\label{spheresS3}
{\rm d}\bar{\Omega}_3^2 = {\rm d}\bar{\theta}^2+\sin^2{\bar{\theta}}
{\rm d}\bar{\phi}^2+\cos^2\bar{\theta} {\rm d}\bar{\psi}^2 \,, \quad
{\rm d} \Omega_3^2 = {\rm d}\theta^2+\sin^2\theta {\rm
  d}\phi^2+\cos^2\theta {\rm d}\psi^2\,.
\end{equation}
Here ($\bar{\theta},\bar{\phi},\bar{\psi}$) parametrise the $S^3$ in
$\text{AdS}_5$, and unbarred angular coordinates parametrise the $S^5$.  In
these coordinates the origin of the $\text{AdS}$ space is at $u=R/2$,
while the boundary is an $S^3$ at $u\rightarrow \infty$.

\begin{figure}[th]
\begin{center}
\includegraphics[width = 0.48\textwidth]{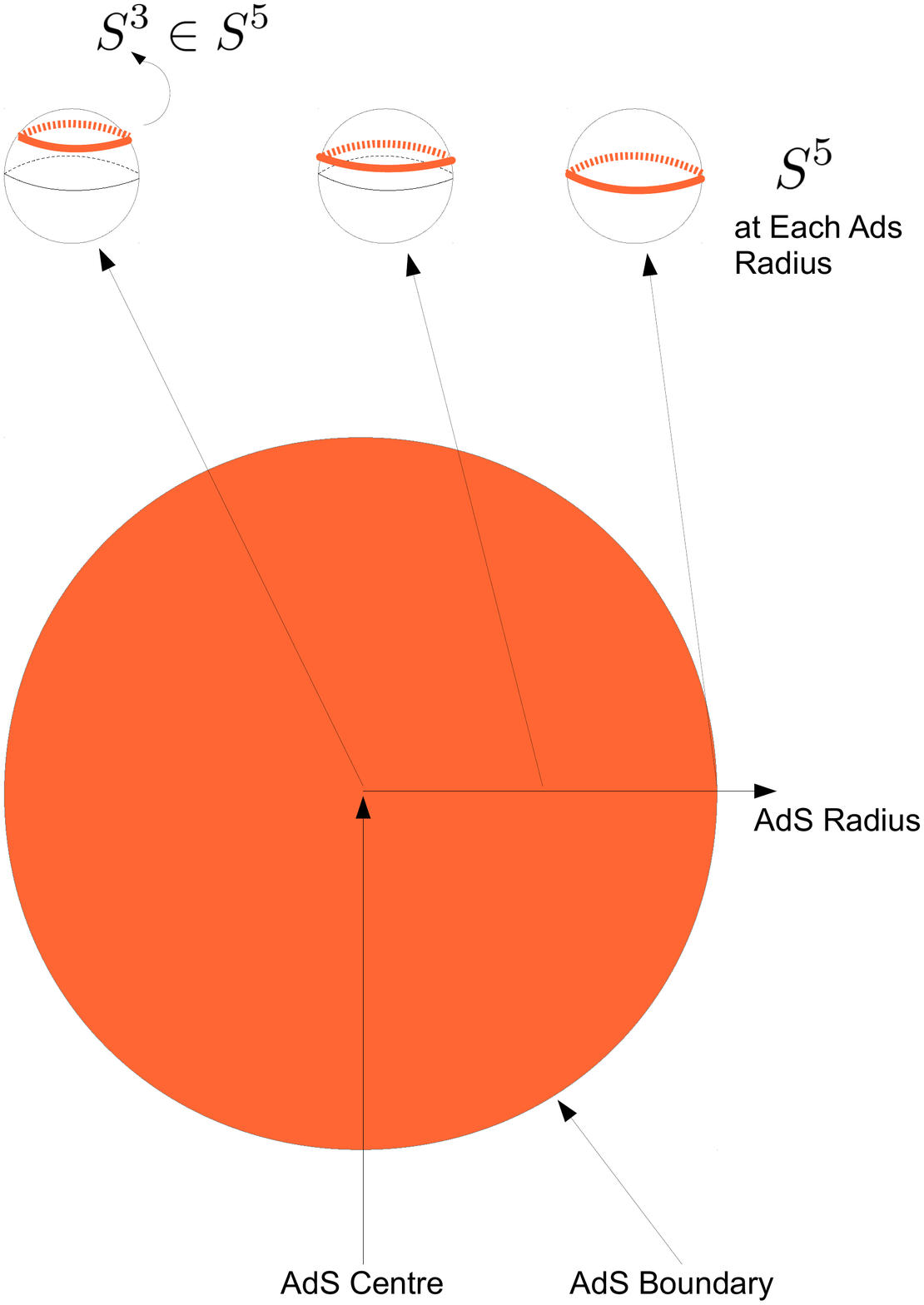}
\includegraphics[width = 0.48\textwidth]{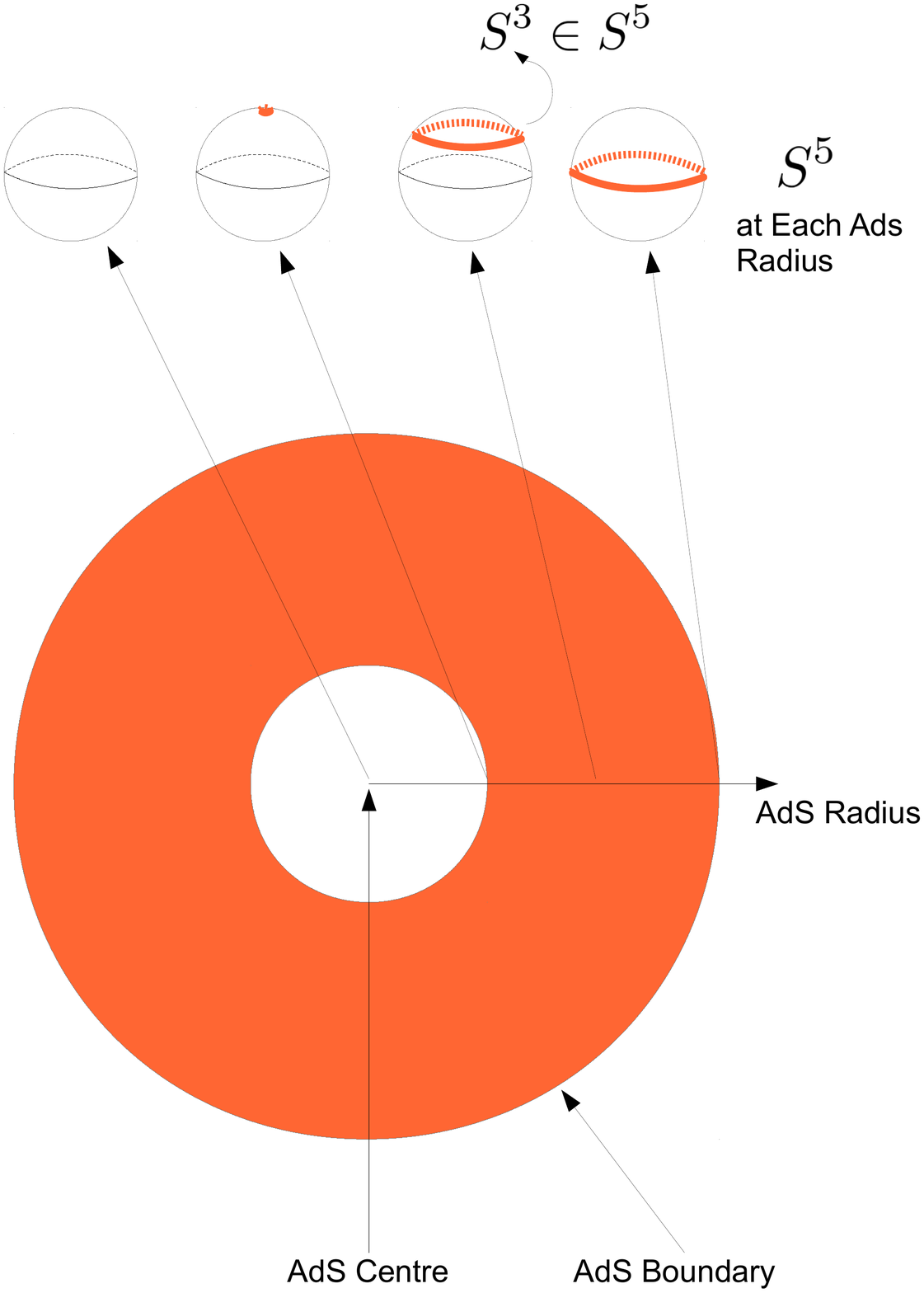}
\end{center}
\caption{Schematic depiction of the two types of D7-brane embeddings in
  global $\text{AdS}_5\times S^5$. The D7-brane is highlighted in
  orange. The panel on the left shows a ball embedding, which
  completely fills the $\text{AdS}_5$ space. The panel on the right
  shows a Minkowski embedding. It shrinks on the $S^5$ at some point
  before it reaches the origin of $\text{AdS}_5$.\label{f:embeddings}}
\end{figure}

In order to reliably do numerical computations it will be useful to
switch to compact coordinates for global AdS space, in which the
metric reads
\begin{equation}
{\rm d}s^2 = 
   - \frac{1}{4z^2} \left(1 + z^2\right)^2 {\rm d}t^2 
   + \frac{R^2}{4z^2} \left(1-z^2 \right)^2 {\rm d}\bar{\Omega}_3^2  
   + R^2 \frac{{\rm d}z^2}{z^2} + R^2 {\rm d}\Omega_5^2 \, .
\end{equation}
The $z$-coordinate related to the non-compact $u$-coordinate by
\begin{equation}
\label{noncompt}
z=\frac{R}{2u} \,.
\end{equation}
Let us note that in the $z$-coordinate the origin of the AdS space is at
$z=1$ while the boundary is at $z=0$.

Introducing D7-probe branes in this geometry corresponds, in the
holographic language, to adding flavour hypermultiplets to
$\mathcal{N}=4$ SYM on the sphere $S^3$. A study of various D-brane
probes, in particular D7-probe branes, in this geometry was performed
in \cite{Karch:2006bv,Karch:2009ph}. It was found that at zero (and
low) temperature, there are two possible D7-brane embeddings in this
dual geometry. The first type of embedding consists of those D7-branes
which completely fill the $\text{AdS}_5$ space.  The second type of
embedding consists of the D7-branes which wrap a non-maximal $S^3\in
S^5$ which shrinks along the radial direction of $\text{AdS}_5$ and
collapses before the brane reaches the origin of $\text{AdS}_5$. See
figure~\ref{f:embeddings} for a schematic comparison.

Within the first series there is one particular embedding in which the
brane wraps the equatorial $S^3 \in S^5$ (i.e.~it is a brane with
vanishing extrinsic curvature). This system is dual to the $\mathcal
N=2$ SYM theory with massless hypermultiplets. All other embeddings in
this class, and all embeddings in the Minkowski class, are dual to a
massive hypermultiplet, with a mass related to the distance at which
the D7-brane ``stops'' before the origin of the
$\text{AdS}_5$\footnote{A more correct description of the mass is of
  course given in terms of asymptotic data.}. Interestingly, as the
``quark'' mass is varied a topology changing phase transition occurs
\cite{Karch:2006bv}. It was further analysed in detail in
\cite{Karch:2009ph} that this phase transition is actually third
order, unlike most of the phase transitions associated to probe branes
in holographic duals, which are usually first order.

In the high-temperature phase, the situation is similar to that in
infinite volume. One finds that Lorentzian and black hole embeddings
exhaust all possibilities. Just as in infinite volume, these
correspond to D7-branes that stay outside the horizon or reach the
black hole horizon respectively. We will not discuss the
high-temperature phase in this paper.

\section{Introducing a magnetic field into the system}

In the previous section we have recalled that empty global $\text{AdS}_5
\times S^5$ admits two types of D7-branes, depending on the quark
mass. The question we want to address now is what happens to the brane
embeddings when we introduce an external magnetic field in the
direction of the boundary $S^3$.

\subsection{Equations of motion and the ansatz}
\label{s:EOM}

The equations of motion for the probe brane with flux follow from the
DBI action
\begin{equation}\label{DBI}
S = -T_{\text{D7}}\int \!{\rm d}^8\sigma\sqrt{-\det(e_{ab}+2\pi\alpha' F_{ab})}\,,
\end{equation}
where $e_{ab} = E_{\mu\nu}\partial_ax^\mu\partial_bx^\nu$ with
$E_{\mu\nu} = G_{\mu\nu}+B_{\mu\nu}.$ We obtain the equations of
motion for the embedding function and for the gauge field,
\begin{gather}
\label{eome}
\partial_b\bigg{(}\sqrt{-{\cal E}}{\cal E}^{ba}E_{\nu\lambda}\frac{\partial x^\nu}{\partial \sigma^a}\bigg{)}+\partial_a\bigg{(} \sqrt{-{\cal E}}{\cal E}^{ba}E_{\lambda\nu}\frac{\partial x^\nu}{\partial \sigma^b} \bigg{)} -\sqrt{-{\cal E}}{\cal E}^{ba}\partial_\lambda E_{\mu\nu}\frac{\partial x^\mu}{\partial \sigma^a}\frac{\partial x^\nu}{\partial \sigma^b} = 0\,,\\[1ex]
\label{eoma}
\partial_a(\sqrt{-{\cal E}}({\cal E}^{ab}-{\cal E}^{ba}))=0\,,
\end{gather}
where ${\cal E}^{ab}$ and ${\cal E}$ are inverse and determinant
of ${\cal E}_{ab} = e_{ab} + 2\pi\alpha'F_{ab},$ respectively.

The action (\ref{DBI}) is invariant under the gauge symmetry which
involves, on equal footing, the Yang-Mills field strength $F_{ab}$ on
the brane worldvolume and the pull-back of the two form gauge
potential $B_{\mu\nu}$ in the bulk. Hence, having a nontrivial magnetic
field strength on the brane worldvolume is equivalent to having
a nontrivial $B$-flux in the bulk, and would in principle lead to
backreaction to the background geometry (even in the probe-brane
limit).  In order to avoid backreaction of the magnetic field on
the D7-brane on the background geometry we will therefore take the
$B$-field to be pure gauge. We make the following ansatz,
\begin{equation}\label{backgroundB}
B = \frac{1}{2} H'(u)R^2(\sin^2\bar{\theta}{\rm d}u{\rm d}\bar{\phi}+\cos^2\bar{\theta}{\rm d}u{\rm d}\bar{\psi})+
H(u)R^2\sin\bar{\theta}\cos\bar\theta({\rm d}\bar\theta {\rm d}\bar\phi-
{\rm d}\bar\theta {\rm d}\bar\psi) \, .
\end{equation}
It is easy to see that the $B$-field is indeed pure gauge, $B = {\rm d} \Lambda$,
where
\begin{equation}
\Lambda = \frac{1}{2} H(u)R^2(\sin^2\bar\theta {\rm d}\bar\phi + \cos^2\bar\theta {\rm d}\bar\psi) \, .
\end{equation}
When writing this ansatz we have used the simplest, lowest lying
vector spherical harmonic on $S^3$, with quantum numbers
$(\bar{l}=1,\bar{s}=-1)$ (or $\bar{l}=1,\bar{s}=1$).  While a linear
combination of the two would also be pure gauge, it does not satisfy
the equations of motion, so in what follows we focus on a single
lowest lying spherical harmonic.

We should emphasise that our ansatz differs from the magnetic field
studied in \cite{Filev:2012ch}. When constructing the ansatz for the
magnetic field, their idea was to used the tetrads of the metric on
the $S^3$ (\ref{spheresS3}), as these being one-forms would provide a
pure gauge magnetic field.  While this is true, as we show below, the
ansatz of \cite{Filev:2012ch} does not satisfy the second equation of
motion (\ref{eoma}) and as such does not describe a physical magnetic
field. The ansatz for the magnetic field used in \cite{Filev:2012ch}
is given by
\begin{equation}
\label{ansatzwrong}
B=He^{(1)}\wedge e^{(2)}=H R^2\sin\bar\theta {\rm d}\bar\theta {\rm d}\bar\phi = {\rm d}(-HR^2\cos\bar\theta {\rm d}\bar\phi)\,,
\end{equation}
where the local tetrads are given by
\begin{equation}
e^{(1)} = R {\rm d}\bar\theta,\qquad
e^{(2)} = R\sin\bar\theta {\rm d}\bar\phi,\qquad
e^{(3)} = R\cos\bar\theta {\rm d}\bar\psi \, .
\end{equation}
In order to show that (\ref{ansatzwrong}) does not satisfy the second
equation of motion (\ref{eoma}), we split the inverse matrix
$\mathcal{E}^{ab}$ into two parts as follows
\begin{equation}
\mathcal{E}^{ab} = S^{ab} + J^{ab}
\end{equation}
where $S^{ab}$ is the symmetric while $J^{ab}$ is the antisymmetric
part of $\mathcal{E}^{ab}$. When evaluated on the ansatz
(\ref{ansatzwrong}) the antisymmetric part $J^{ab}$ becomes
\begin{equation}
J^{\bar\theta\bar\phi}=-\frac{H R^2}{\sin\bar\theta\left( H^2 R^4 + \left( u-\frac{R^2}{4 u}\right)^4 \right)} = -J^{\bar\phi\bar\theta}.
\end{equation}
and hence the left hand side of the second equation of motion
(\ref{eoma}) becomes
\begin{equation}
\begin{aligned}
\partial_a(\sqrt{-\mathcal{E}}(\mathcal{E}^{a\bar\phi}- \mathcal{E}^{\bar\phi a}))&=\partial_{\bar\theta}(\sqrt{- \mathcal{E}}(\mathcal{E}^{\bar\theta\bar\phi}- \mathcal{E}^{\bar\phi\bar\theta})) \\
	&=-\frac{2H R^2}{H^2 R^4+\left(u-\frac{R^2}{4 u}\right)^4}\partial_{\bar\theta}\left(\frac{\sqrt{-\mathcal{E}}}{\sin\bar\theta}\right) \\
	&=\frac{2 H R^5 u}{\sqrt{H^2 R^4+\left(u-\frac{R^2}{4 u}\right)^4}}\sin\bar\theta\sin\theta\cos\theta \cos^3(\theta_3(u)) \\
	&\qquad\qquad \times\left(1-\frac{R^4}{16 u^4}\right)\sqrt{1+u^2\theta_3'(u)^2} \\
	&\neq 0 \,  .
\end{aligned}
\end{equation}
Hence in what follows, we focus on the ansatz \eqref{backgroundB} for
the magnetic field which is based on spherical harmonics, and can be
made to satisfy all equations of motion. As we will see the physical
solutions for the brane embeddings are qualitatively very different
from the results which were obtained in \cite{Filev:2012ch}.

We also need to make an ansatz for the brane embedding. Since the
magnetic field is turned on in the direction of the boundary $S^3$, we
only need one transverse scalar to describe the shape of the
D7-brane. While the external magnetic field introduced via
\eqref{backgroundB} on the boundary sphere is obviously not constant,
its norm, however, is. Moreover, it can be shown that our ansatz
preserves an $SO(3)\times SO(2)$ subgroup of the full $SO(4)$ isometry
group of the boundary $S^3$. Assuming that the scalar $\chi$ preserves
the same amount of symmetry, this then implies that it can
depend only on the holographic direction $u$. Hence we make the
ansatz for the scalar which field which depends only on the
$u$-coordinate,
\begin{equation}\label{em2}
\chi = \chi(u)\,,\quad \kappa = 0.
\end{equation}
Substituting the ansatz \eqref{backgroundB} and \eqref{em2} into the
action one gets
\begin{multline}
S = -T_{\text{D7}}\frac{R^3}{2}\int\! {\rm d}^8\sigma\sin(\theta)\cos(\theta)\sin(\bar\theta)\cos(\bar\theta)\frac{u}{R}\left(1+\frac{R^2}{4u^2}\right)  \\[1ex]
\times \sqrt{\left(\left(1-\chi (u)^2\right) \left(4 \left(1-\frac{R^2}{4u^2}\right)^2 R^2+H'(u)^2R^4\right)+4 u^2\left(1-\frac{R^2}{4u^2}\right)^2R^2 \chi '(u)^2\right)} \\[1ex]
\times\sqrt{\left(u^4\left(1-\frac{R^2}{4u^2}\right)^4+H(u)^2R^4\right)}\left(1-\chi (u)^2\right)\,.
\end{multline}
The equation of motion for $\chi$ that follows from this action is given by
\begin{align}
\label{eomchi}
0&= \chi ''(u) -\frac{4 \chi (u) \chi '(u)^2}{\chi (u)^2-1}  + \frac{3 \chi (u) \left(4 R^2 u^4 H'(u)^2+\left(R^2-4 u^2\right)^2\right)}{\left(R^2 u-4 u^3\right)^2}  \\
&+\frac{\left(256 R^8 u^4 H(u)^2+12288 R^4 u^8 H(u)^2+\left(R^2-4 u^2\right)^4 \left(3 R^4+16 R^2 u^2+80 u^4\right)\right) \chi '(u)}{u \left(4 u^2-R^2\right) \left(R^2+4 u^2\right) \left(256 R^4 u^4 H(u)^2+\left(R^2-4 u^2\right)^4\right)} \nonumber \\
&+\frac{\left(1024 R^6 u^8 H(u)^2+4 \left(R^2-4 u^2\right)^2 \left(3 R^6 u^4+8 R^4 u^6+48 R^2 u^8\right)\right) H'(u)^2 \chi '(u)}{u \left(4 u^2-R^2\right) \left(R^2+4 u^2\right) \left(256 R^4 u^4 H(u)^2+\left(R^2-4 u^2\right)^4\right)} \nonumber\\
&+\frac{4 u \left(128 R^4 u^4 H(u)^2 \left(R^4+16 u^4\right)+\left(R^2-4 u^2\right)^4 \left(R^4+4 R^2 u^2+16 u^4\right)\right) \chi '(u)^3}{\left(R^2-4 u^2\right) \left(R^2+4 u^2\right) \left(\chi (u)^2-1\right) \left(256 R^4 u^4 H(u)^2+\left(R^2-4 u^2\right)^4\right)} \,.\nonumber
\end{align}
\smallskip

\noindent The equation of motion for the $B$-field, expressed in terms of the
flux $H$, reads

\begin{align}
\label{eomH}
0&=H''(u) -\frac{256 R^4 u^4 H(u) H'(u)^2}{256 R^4 u^4 H(u)^2+\left(R^2-4 u^2\right)^4}\nonumber \\
&-\frac{4 R^2 u^3 \left(256 R^4 u^4 H(u)^2+\left(R^2-4 u^2\right)^2 \left(3 R^4+8 R^2 u^2+48 u^4\right)\right) H'(u)^3}{\left(R^2-4 u^2\right) \left(R^2+4 u^2\right) \left(256 R^4 u^4 H(u)^2+\left(R^2-4 u^2\right)^4\right)}\nonumber \\
&+ \frac{64 H(u) \left(R^3-4 R u^2\right)^2 \left(u^2 \chi '(u)^2-\chi (u)^2+1\right)}{\left(\chi (u)^2-1\right) \left(256 R^4 u^4 H(u)^2+\left(R^2-4 u^2\right)^4\right)}\\
&+ \frac{\left(\left(R^2-4 u^2\right)^4 \left(R^4+48 u^4\right)-256 R^4 u^4 H(u)^2 \left(R^4+16 R^2 u^2-16 u^4\right)\right) H'(u)}{u \left(4 u^2-R^2\right) \left(R^2+4 u^2\right) \left(256 R^4 u^4 H(u)^2+\left(R^2-4 u^2\right)^4\right)} \nonumber  \\
&+\frac{4 u \left(128 R^4 u^4 H(u)^2 \left(R^4+16 u^4\right)+\left(R^2-4 u^2\right)^4 \left(R^4+4 R^2 u^2+16 u^4\right)\right) H'(u) \chi '(u)^2}{\left(R^2-4 u^2\right) \left(R^2+4 u^2\right) \left(\chi (u)^2-1\right) \left(256 R^4 u^4 H(u)^2+\left(R^2-4 u^2\right)^4\right)}\,.\nonumber
\end{align}
These same equations can be obtained by first varying the action and
then inserting the ansatz, as we have explicitly verified.

\section{Solving the equations of motion}

We now want to solve the equations of motion (\ref{eomH}), (\ref{eomchi})
for the brane shape and worldvolume flux. As reviewed before, in the
absence of flux there are two possible embeddings of the D7-brane,
Minkowski and ball embeddings. We expect that these two classes will
continue to exist, at least for small values of the magnetic field. We
therefore impose the same type of boundary conditions on the brane
shape as in the absence of the worldvolume flux, and then solve the equations
including the magnetic field.

For Minkowski embeddings, we impose that the D7-brane caps off at some
finite distance (IR ``endpoint'') $u=u_{\text{Mink}}$.  In other
words, the $S^3$ in $S^5$ wrapped by the brane shrinks to zero size,
so that $\chi(u=u_{\text{Mink}}) = 1$. In addition we require that
there is no conical singularity present at the tip of the brane (as in
\cite{Karch:2006bv}). This implies that the general expansion of the
D7-brane near the tip is given by
\begin{equation}
\label{expansionMink}
\begin{aligned}
\chi(u) &= 1+\chi_{M1}(u-u_{\text{Mink}}) + \chi_{M2}(u-u_{\text{Mink}})^2 + O((u-u_{\text{Mink}})^3) \,, \\[1ex]
H(u) &= H_M + H_{M1}(u-u_{\text{Mink}}) + H_{M2}(u-u_{\text{Mink}})^2 + O((u-u_{\text{Mink}})^3) \, , 
\end{aligned}
\end{equation}
where $\chi_{M1},H_{M1}$ are long expressions of $u_{\text{Mink}}, H_M.$

For the ball embeddings, the D7-brane reaches all the way to the
$\text{AdS}$ origin where the $S^3$ in $AdS_5$ shrinks to zero size,
while the $S^3$ in $S^5$ remains of finite size at that point. Since
the magnetic field points in the direction of the shrinking $S^3$ we
have to require that at the origin of the $\text{AdS}_5$ space the
magnetic field vanishes. We also impose a Neumann boundary condition
for the brane shape at the origin of the AdS space
$\chi'(u=u_{\text{ball}})=0$, again similar to
\cite{Karch:2006bv}. Putting all this together, we have for the
general expansion of the ball brane near the AdS origin
\begin{equation}
\label{expansionBall}
\begin{aligned}
\chi(u) &= \chi_{\text{ball}}-\frac{3}{16R^2}
\left(-\frac{R}{2}+u\right)^2 \left(16 +H_{c2}^2 \right)\chi_{\text{ball}} +
O((u-R/2)^3)  \, ,  \\[1ex]
H(u) &= \frac{H_{c2}}{R^2}
\left(-\frac{R}{2}+u\right)^2-2 \frac{H_{c2}}{R^3}
\left(-\frac{R}{2}+u\right)^3 + O((u-R/2)^4) \, ,
\end{aligned}
\end{equation}
where $\chi_{\text{ball}}$ parametrises the size of the $S^3$ in $S^5$ at the
origin of the AdS space.

We also need the expansion of the scalar $\chi$ and magnetic field $H$
at infinity, in order to read off the physical parameters of the
system. The general expansion of the bulk fields near the AdS boundary is
given by
\begin{equation}
\begin{aligned}
\chi(u) &= \frac{m}{u} + \frac{c_1}{u^3} - \frac{m R^2}{2u^3}\log \frac{u}{R} + O(u^{-4})\,,\\[1ex]
H(u) &= H_{\text{ext}} + \frac{M_{tz}}{u^2} - \frac{2H_{\text{ext}}R^2}{u^2}\log \frac{u}{R} + O(u^{-3})\,.
\end{aligned}
\end{equation}
In order to get rid of the divergent logarithmic terms, we need to
renormalise the parameters in this expansion, following the
holographic renormalisation procedure which was developed
in~\cite{Karch:2005ms}.  Full details of this derivation are presented
in~\cite{Vanichchapongjaroen:2014a}.  One finds for the renormalised
parameters
\begin{equation}\label{ren}
c = c_1 - \frac{mR^2}{2}\log\frac{m}{R}\,,
\quad M = M_{tz} + H_{\text{ext}}R^2\,.
\end{equation}
The physical parameters of the system are the bare ``quark'' mass
$m_q=m/2\pi \alpha'$ (i.e.~the mass of the fermions in the hypermultiplet)
and its condensate $c$. As the system is exposed to the nonconstant
external magnetic field (\ref{backgroundB}) with ``amplitude''
$H_{\text{ext}}$, a non-vanishing magnetisation $M$ is present in the ground
state.  Our goal now is to understand the geometry of the branes as a function
of the external magnetic field, and to determine how the magnetisation and the
condensate vary as the external field is varied.

The equations of motion for the scalar $\chi$ and magnetic field $H$
are complicated, and in order to solve them we need to resort to
numerical methods. We solve the equations directly by starting the
numerical evolution from the origin (for ball branes) or the tip of
the D7-brane (for Minkowski branes), as
in~\cite{Karch:2006bv}. However, as a cross check of our results, we
have also used the more complicated shooting method which starts from
the boundary of the AdS space and scans through the parameter space in
order to find solutions that satisfy the boundary conditions at the
origin or tip. While it is possible to perform a full numerical
analysis in the $u$-coordinates (and we have solved the equations in
this system as well), in order to have a numerical solution which
covers the full AdS space (with no asymptotic cutoff), we present here
the results in the compact coordinates~\eqref{noncompt}, in which the
origin of the AdS space is at $z=1$ and the boundary is at $z=0$.

As said above, integrating outwards from the origin or tip of AdS is
much simpler than shooting from the boundary. The main point is that
since the regularity conditions for the behaviour of the brane at the
origin or tip have already been built in to the
expansions~\eqref{expansionBall}, \eqref{expansionMink}, one always
finds a physical solution for any value of the two free parameters
(the position of the brane at the origin or tip, and the value of the
magnetic field there).  On the other hand, when shooting from
infinity, one has to scan through a four dimensional space of
parameters while searching for a small two-dimensional subspace of
solutions which behave in a regular fashion near the origin. The price
one has to pay, however, when shooting from the origin is that,
although the computation is simpler, the physical parameters can only
be read off \emph{after} the solutions have been constructed. So if
one encounters a situation, as we will, that there appear forbidden
regions in the physical parameter space, one cannot easily see what
goes ``wrong'' with the solutions in this region of parameter
space. For this reason we considered it essential to be able to
reproduce the results using these two different methods, integrating
from the origin/tip or shooting from the boundary. Some more technical
comments on the techniques used to shoot from the boundary will be
presented in section~\ref{s:numerics}.

\begin{figure}[th]
\begin{center}
\vspace{2ex}
\includegraphics[width = 0.48\textwidth]{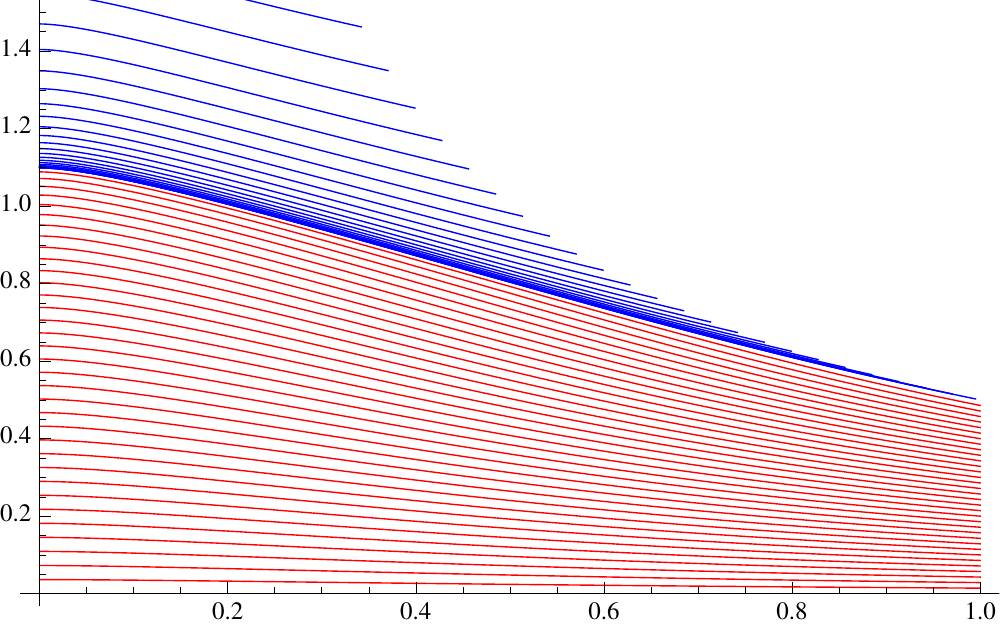}
\includegraphics[width = 0.48\textwidth]{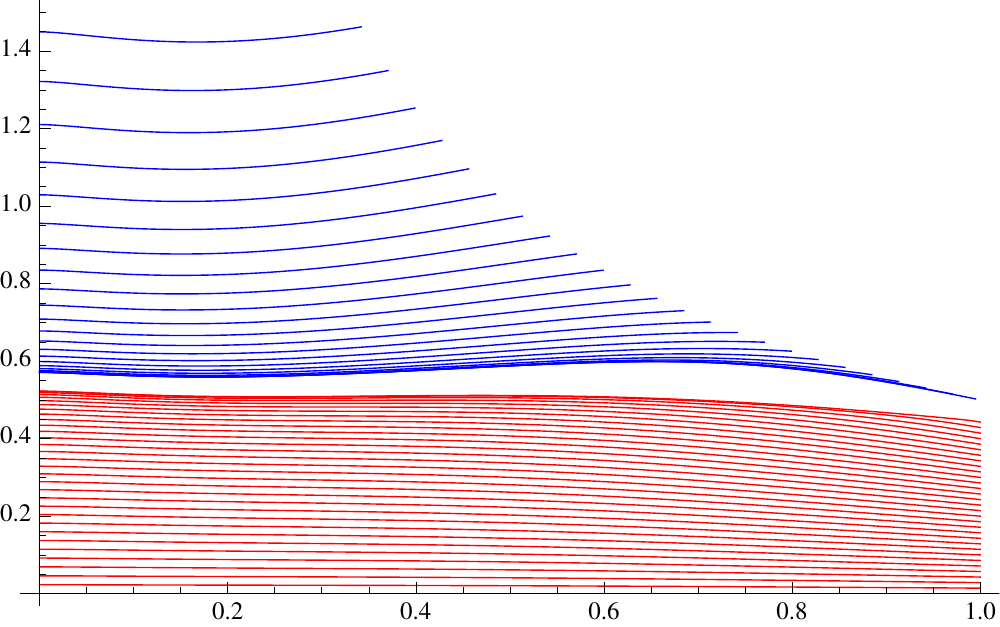}\\[2ex]
\includegraphics[width = 0.48\textwidth]{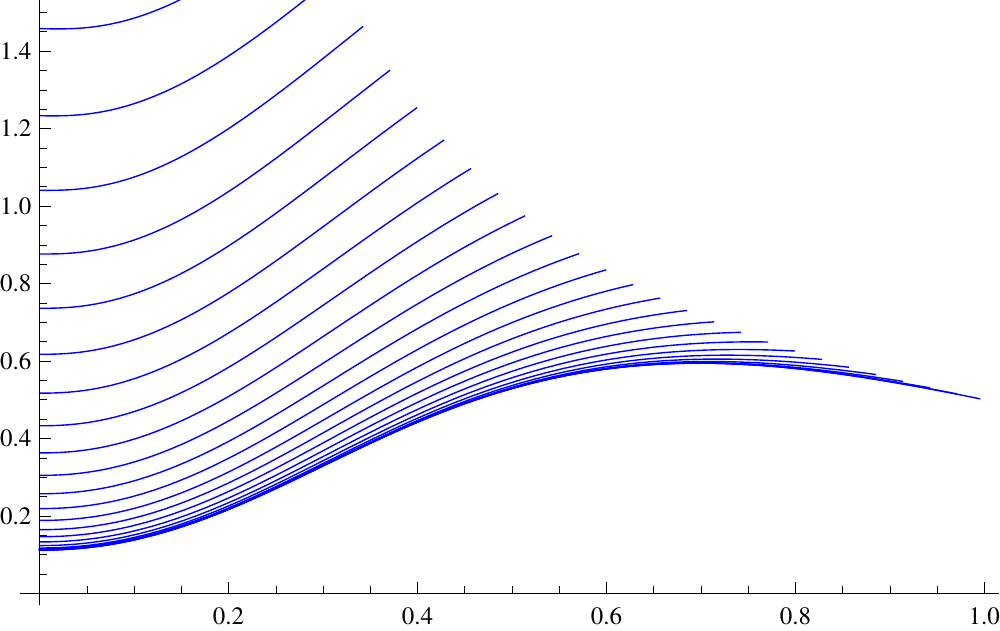}
\includegraphics[width = 0.48\textwidth]{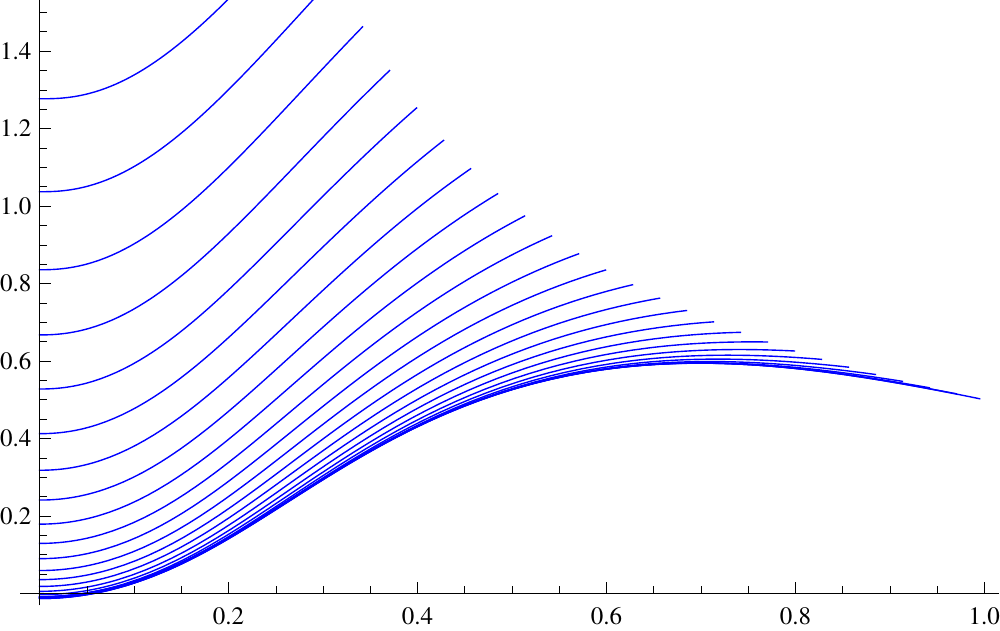}
\end{center}
\caption{Brane shapes for D7 embeddings in the $z$-coordinate.  From
  left to right and top to bottom the external magnetic field takes
  values $H_{\text{ext}} = 0,2,4,5$ respectively.  Red is for ball embeddings
  where D7 reaches the AdS centre, while blue is for Minkowski
  embeddings where the D7 does not reach the AdS centre.  The
  horizontal and vertical axes label $z$ and $\chi/(2z)$. The quark
  mass of each embedding is proportional to the position of the brane
  on the vertical axis.  The plots clearly show the development of a
  mass gap.\label{figures1}}
\end{figure}

\begin{figure}[th]
\vspace{.2cm}
\begin{center}
\includegraphics[width = 0.3\textwidth]{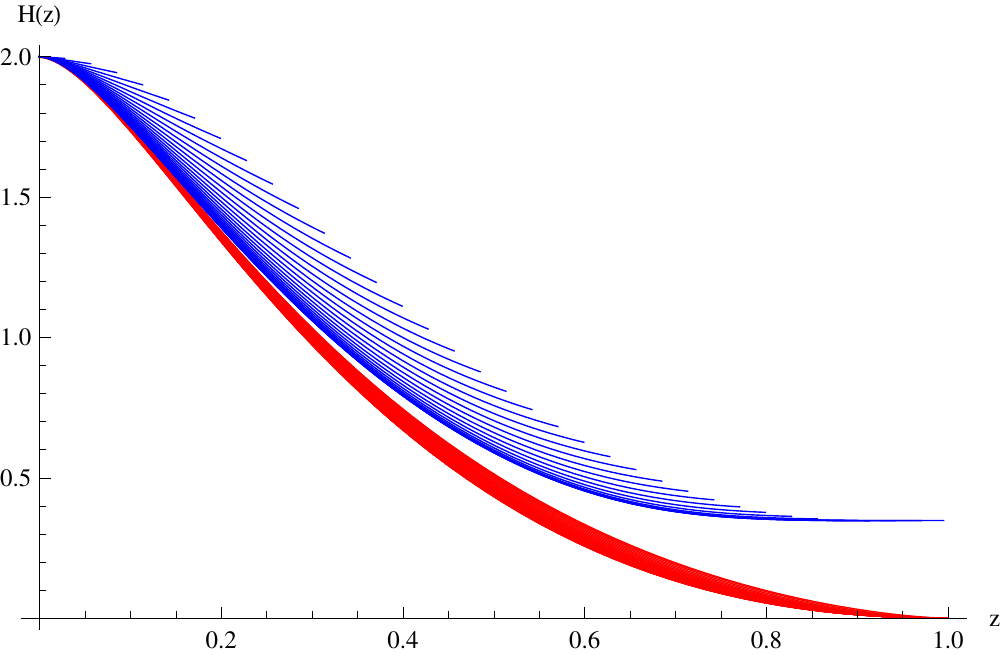}
\includegraphics[width = 0.3\textwidth]{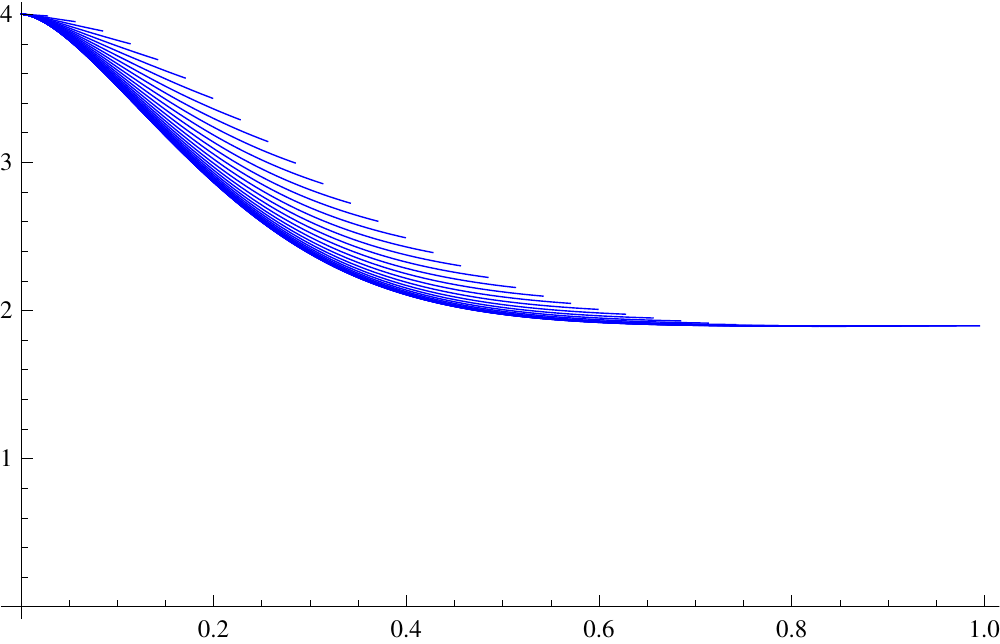}
\includegraphics[width = 0.3\textwidth]{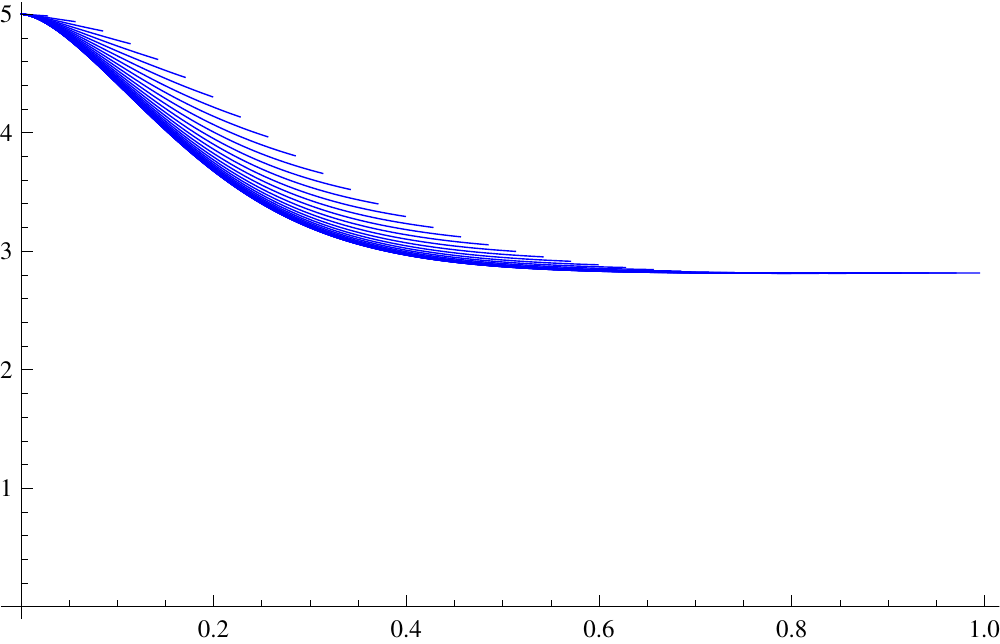}
\end{center}
\vspace{-.5cm}
\caption{Plot of the profile $H(z)$ for various brane embeddings.
  From left to right the external magnetic field takes values
  $H_{\text{ext}} = 2,4,5$ respectively.  Red is for ball embeddings
  where the D7-brane reaches the AdS centre.  Blue is for Minkowski
  embeddings where the D7-brane does not reach the AdS centre.
\label{figures2}
}
\end{figure}

\begin{figure}[th]
\begin{center}
\includegraphics[width = 0.28\textwidth]{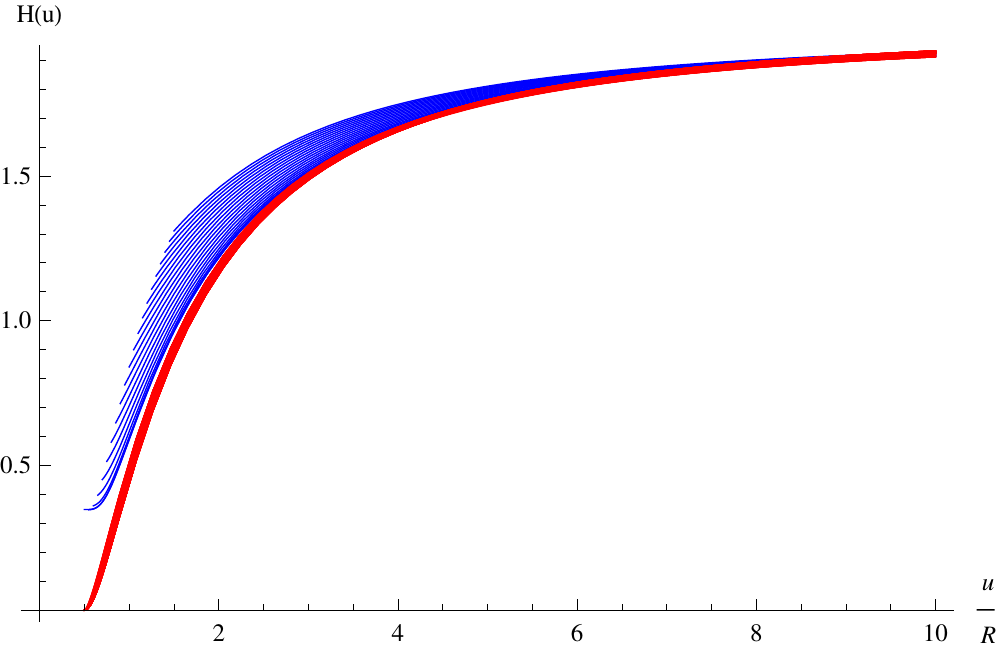}
\includegraphics[width = 0.28\textwidth]{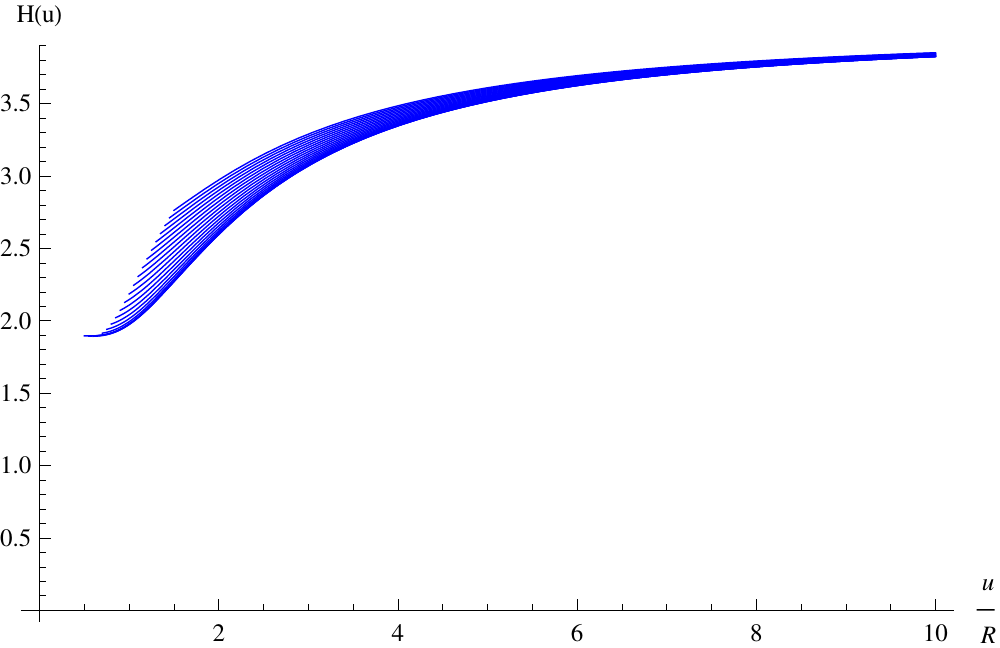}
\includegraphics[width = 0.28\textwidth]{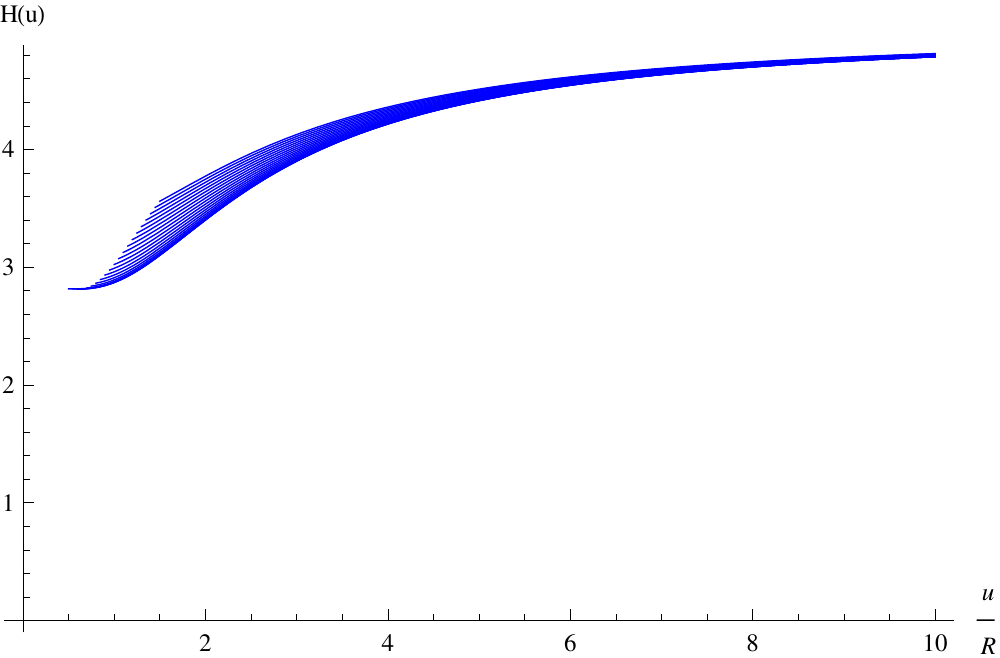}
\end{center}
\caption{Plot of the profile $H(u)$ for different
branes embedding.
From left to right and top to bottom shows
the plot at fixed external magnetic field
$H_{\text{ext}} = 2,4,5,$ respectively.
Red is for ball embedding where D7 reaches AdS centre.
Blue is for Minkowski embedding where D7 does not
reach AdS centre.
\label{figures3}}
\end{figure}

\begin{figure}[th]
\vspace{6ex}
\begin{center}
\includegraphics[width = 0.48\textwidth]{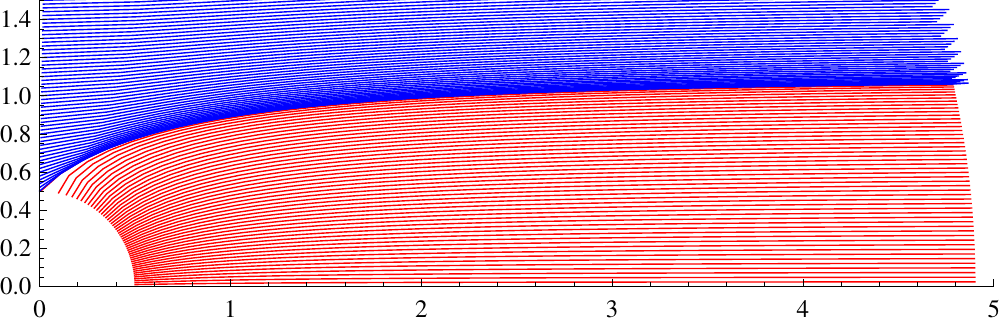}
\includegraphics[width = 0.48\textwidth]{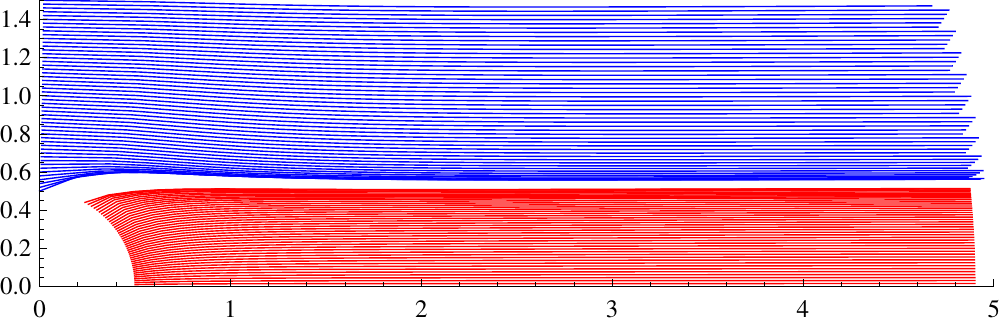}\\[2ex]
\includegraphics[width = 0.48\textwidth]{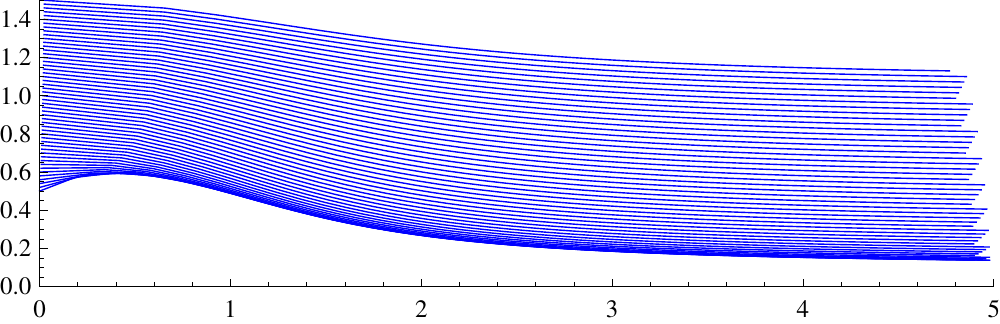}
\includegraphics[width = 0.48\textwidth]{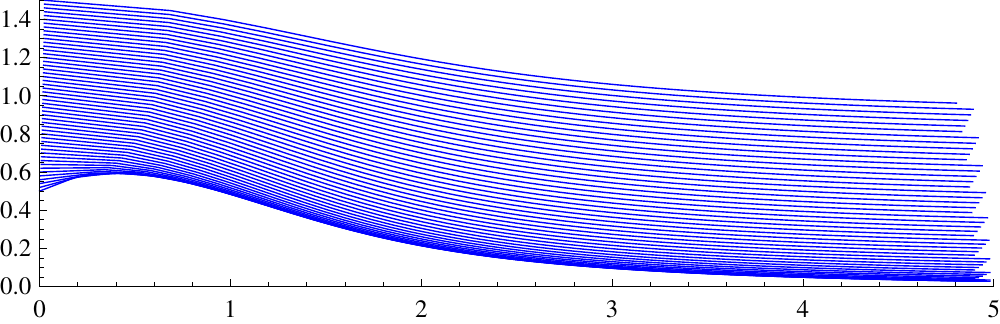}
\end{center}
\caption{Plot of brane shapes for D7 embedding.
From Left to Right and Top to Bottom shows
the plot at fixed external magnetic field
$H_{\text{ext}} = 0,2,4,5,$ respectively.
Red is for Ball embedding where D7 reaches AdS centre.
Blue is for Minkowski embedding where D7 does not
reach AdS centre.
The horizontal and vertical axes are $u\sqrt{1-\chi^2},u\chi.$
We see the development of mass gap as external
magnetic field is increased roughly to $H_{\text{ext}} = 4.$
The gap closes again around $H_{\text{ext}} = 5$
where the lowest Minkowski embedding has
zero quark mass.
}
\label{figures4}
\end{figure}

Solutions for the various shapes of D7-branes are plotted in
figure~\ref{figures1}. We also present the profile of the magnetic
field for various values of the external parameter $H_{\text{ext}}$,
see figure~\ref{figures2}. For completeness, we also present the
shapes and worldvolume magnetic field for various brane embeddings in
the more familiar $u$-coordinates, see figures~\ref{figures3}
and~\ref{figures4}. However, note that in these coordinates one cannot
numerically reach infinity, i.e.~the boundary of AdS space, so all
numerical computations are performed with some cutoff on the
$u$-coordinate.


From the first plot in figure~\ref{figures1} we see that in the
absence of the magnetic field both types of branes are ``pulled'' by
gravity towards the origin of AdS space, i.e.~$z=1$. As the magnetic
field is introduced, the probe branes first tend to be more flat
(i.e.~they fall less in the $\chi$ direction) and then they bend away
in the $\chi$ direction as they get more and more ``repelled'' by the
magnetic field.  In other words, the magnetic field has an effect
opposite from that of gravity.  A similar kind of behaviour was
observed before in the AdS space in Poincar\'e coordinates
in~\cite{Erdmenger:2007bn,Filev:2007gb}.

\begin{figure}[t]
\begin{center}
\includegraphics[width = 0.48\textwidth]{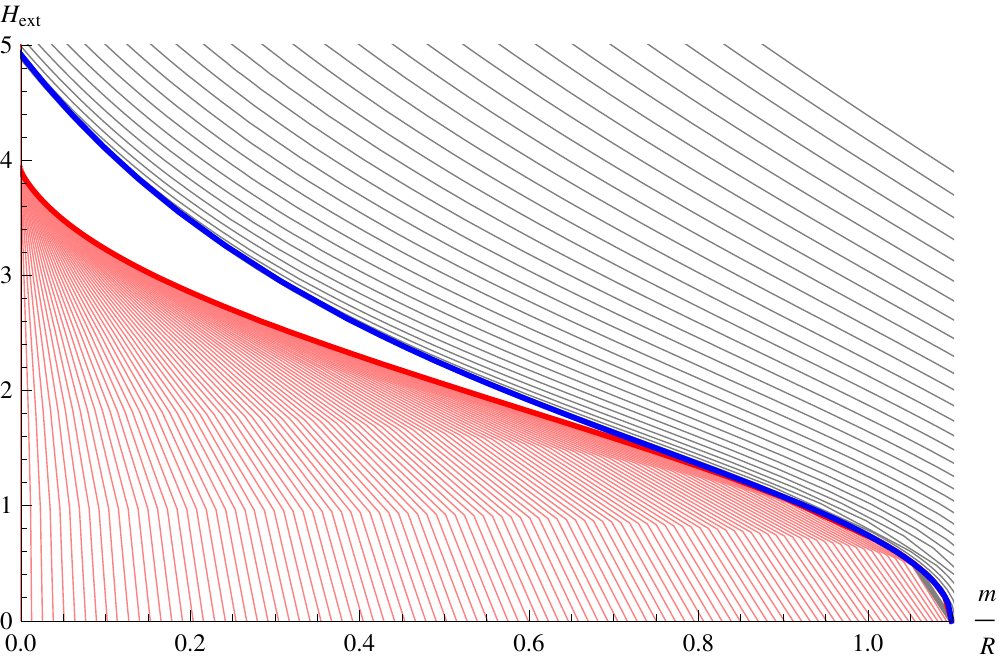}
\includegraphics[width = 0.48\textwidth]{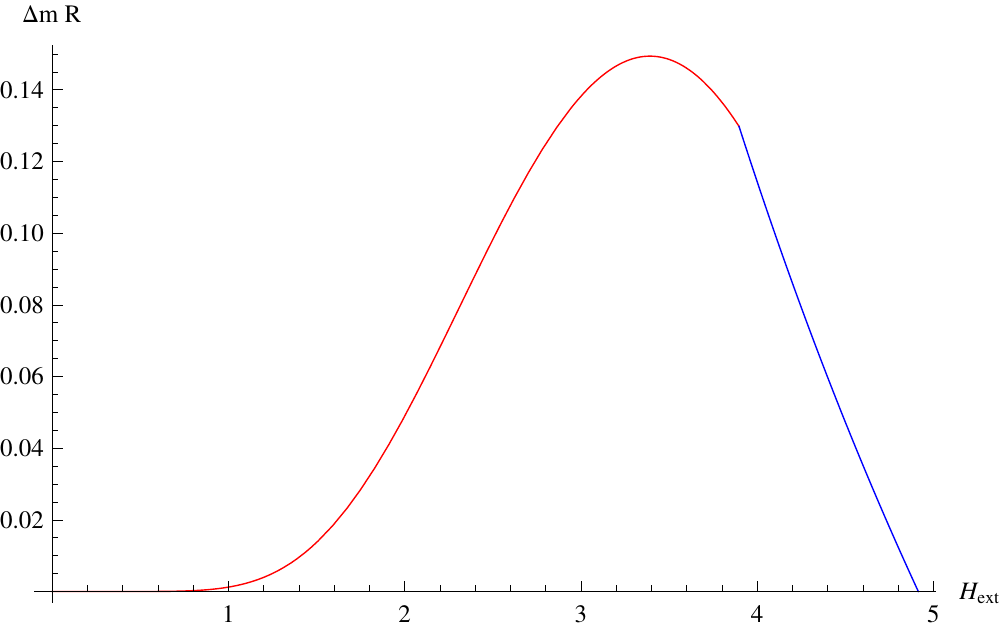}
\end{center}
\caption{The panel on the left shows the possible values of
  $H_{\text{ext}}$ and $m$. The red region corresponds to ball branes,
  while the grey region (above the the blue line) corresponds to
  Minkowski branes, which are all uniquely specified give the values
  $(H_{\text{ext}},m)$.  The curves are just an artifact of the
  discrete steps which were used to scan the full parameter space. The
  panel on the right shows the behaviour of the ``mass gap'' in the
  full parameter space as a function of the external magnetic
  field. The gap begins to develop as soon as $H_{\text{ext}}>0$ and
  it closes again at $H_{\text{ext}} = \hat{H}_2 \approx 5$.}
\label{externalgap}
\end{figure}

We also observe that as the magnetic field is increased, ball branes
characterised by higher mass will start to disappear from the
spectrum. In other words, if the bare quark mass is larger than some
critical value $\Delta m$, then such ball branes cannot exist any
longer (see figure~\ref{externalgap}).  As a consequence, we see that
a ``gap'' between the Minkowski branes phase (upper, blue) and ball
brane phase (lower, red) starts to form. The $y$-axis on these plots
is $\chi/(2z)$, i.e.~it is proportional to the bare quark mass of the
theory, while the horizontal axis labels~$z$. We thus see that the gap
between the two phases is characterised by a gap in the spectrum of
\emph{bare} quark masses.

As the magnetic field is increased, more and more ball branes
disappear from the spectrum. When magnetic field reaches the ``critical''
value $\hat{H}_1 \sim 3.98$, all ball branes have disappeared from the
spectrum and only Minkowski branes are present in the spectrum.

At the same time, the effect of the magnetic field on the Minkowski
branes is to reduce the value for which the Minkowski brane with the
smallest quark mass exists. As the ball branes disappear from the
spectrum, at the same time the spectrum of the Minkowski branes is shifted
towards lower values of the bare quark mass.
As one increases the value of the magnetic field above $\hat{H}_{1}$,
no ball branes are left in the spectrum, and the size of the gap
(i.e.~the non-allowed region for the mass) starts to decrease. When
the magnetic field reaches $\hat{H}_2 \sim 5$, the gap closes, and for
$H > \hat{H}_2$ there is no gap. At this stage even the $m=0$ brane is
Minkowski, and will not reach the origin of the AdS space; this thus
spontaneously breaks the reflection symmetry of the space. The
behaviour of the gap as a function of the external field is shown on
the right hand side of figure~\ref{externalgap}.  This is in 
qualitative agreement with the observation of~\cite{Erdmenger:2007bn}, who
have found, for embeddings in the background of a black hole in the
Poincar\'e patch, that for large enough values of the external magnetic
field no branes can fall into the black hole.

\begin{figure}[th]
\vspace{4ex}
\includegraphics[width = 0.48\textwidth]{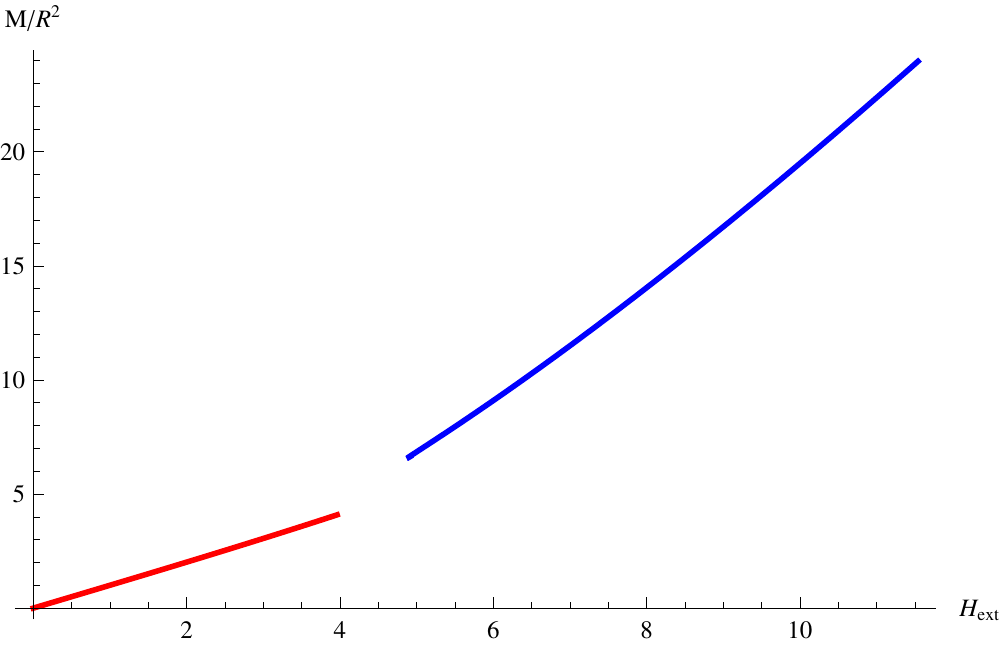}
\includegraphics[width = 0.48\textwidth]{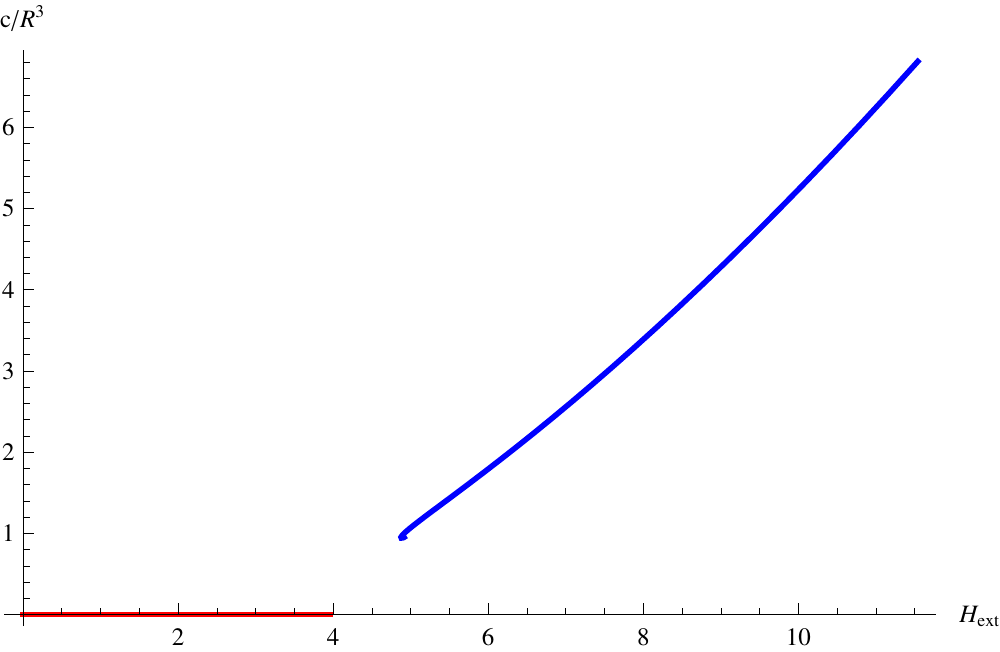}
\caption{The behaviour of the ``condensates'' $M(H_{\text{ext}})$ and
  $c(H_{\text{ext}})$ for $m=0$ branes; see~\eqref{ren} for their
  definition in terms of the bare parameters $c_1$ and $M_{tz}$.}
\end{figure}

\subsection{Comments on the numerical analysis}
\label{s:numerics}

Constructing solutions by starting from the origin is straightforward.
In order to ensure that the mass gap, which we first observed in the brane
embeddings obtained with this method, is not due to an error in our
Mathematica code, we have independently verified these solutions with
an ODE solver built using \cite{Ahnert:2011a}. This revealed no
significant differences.

Constructing solutions by starting from the boundary at $z=0$ is
considerably more challenging, as was already alluded to earlier. At
the boundary we have four parameters at our disposal: $m$,
$H_{\text{ext}}$ and $c_1$ and $M_{tz}$. Since we are mostly interested
in understanding the gap in $m$ for a fixed value of $H_{\text{ext}}$,
we proceed by fixing both these parameters and performing a
two-dimensional search for solutions in the plane spanned by $c_1$ and
$M_{tz}$. One expects on general grounds that there will be a region
in this plane for which the solutions $\chi(z)$ and $H(z)$ remain
regular. Inside that region one then expects two curves, corresponding
to solutions that satisfy one of the two boundary conditions at the
origin (for ball branes) or at the turning point (for Minkowski
branes). Physical solutions lie on the intersection point of these two
curves (if it exists).  However, an explicit analysis shows that the
region of regular solutions is in fact already essentially (up to
numerical errors) one-dimensional.

The two-dimensional scan for regular solutions is done with a
multi-grid method. We start with a grid which, somewhat arbitrarily,
spans $-10\leq c_1/R^3 \leq 10$ and $-10\leq M_{tz}/R^2 \leq 10$ with
unit grid size in both parameters. We label the points on this grid by
a number which describes how close the corresponding solutions are to
being regular (described in more detail below). After this, we
re-initialise the algorithm with a new and denser grid that contains a
set of points for which this number is smallest. Interestingly we find
that the set of regular solutions is essentially one-dimensional, and
all solutions on this curve also satisfy one of the boundary
conditions.  Once we have found this curve, we then switch to an
algorithm somewhat similar to the one in~\cite{BallonBayona:2012wx} to
trace along the curve of regular solutions in the two-dimensional
parameter space. We then either find a point on the curve where both
boundary conditions hold, or no such point (and no physical solution).
\begin{figure}[t]
\includegraphics[height = 7cm]{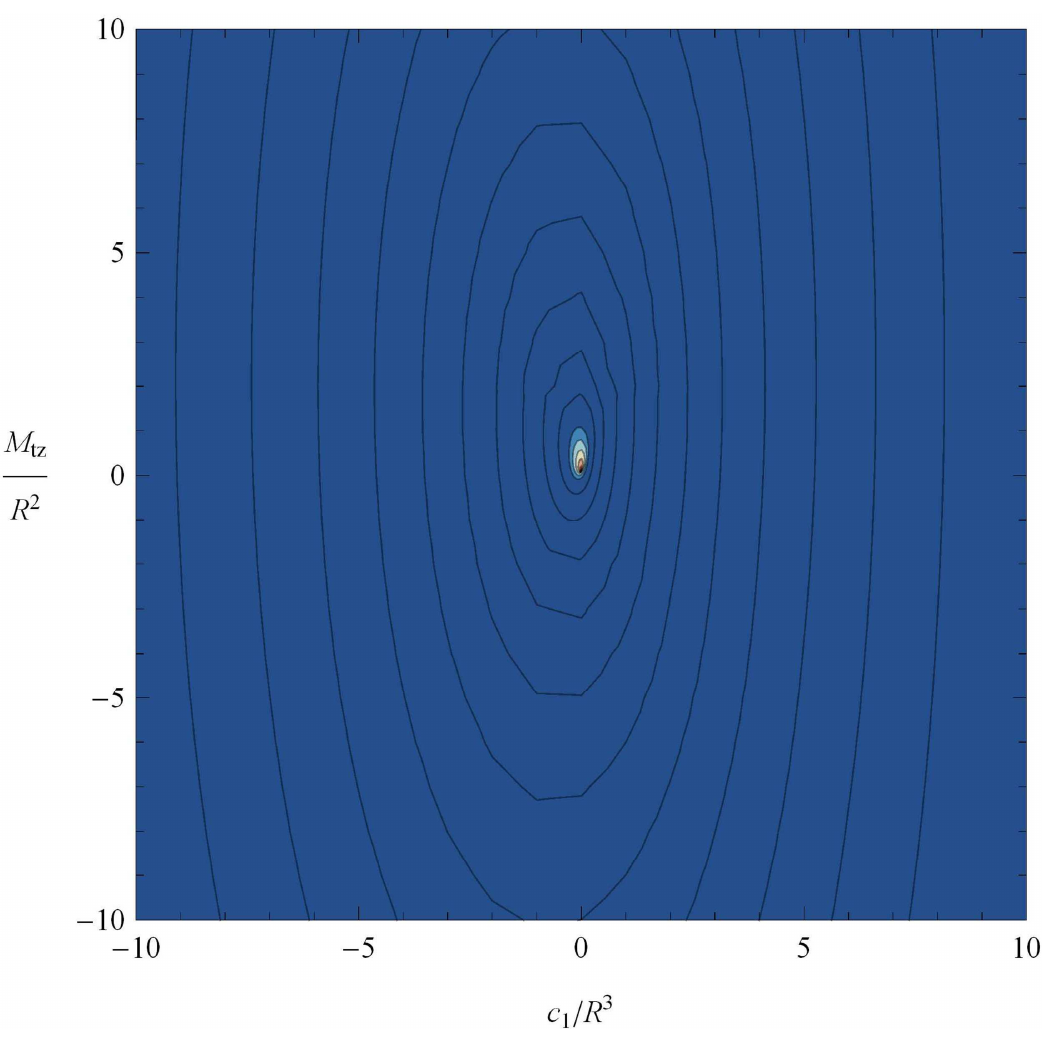}\qquad
\includegraphics[height = 7cm]{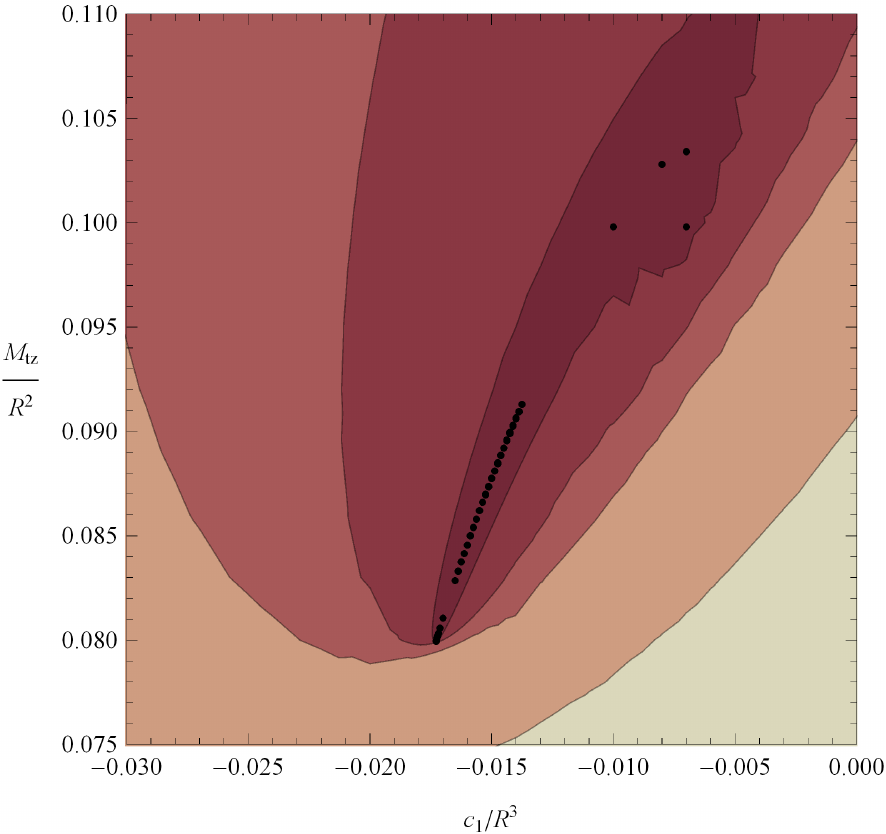}
\vspace{-1ex}
\caption{Typical search for a ball brane solution at
  $(m/R,H_{\text{ext}})=(0.4,1)$, which is known to exist from the
  shooting procedure from the origin (see the left panel of
  figure~\ref{externalgap}). The left plot shows a steep well of
  solutions which get close to a regular solution only in the
  neighbourhood of $(c_1/R^3, M_{tz}/R^2)\approx(-0.015,0.09)$. The
  right plot shows a zoomed region, with black dots representing
  regular solutions, which also satisfy $\theta'(z=1)=0$. The point at
  the bottom end of this curve has $H(z=1)=0$. Red and blue colours
  represent small and large values of the deviation from regularity,
  respectively.\label{f:frombdy_ball}}
\end{figure}

When we are looking for ball solutions, the measure that describes how
close we are to a physical brane solution is taken to be the distance
to the boundary (in the $z$ coordinate) at which either $H'(z)$ or
$\theta'(z)$ diverges. If the solution remains smooth and reaches the
origin, this distance is zero. Such solutions can still have
$H(z=1)\not=0$, so we then we trace the curve of physical solutions
and determine the value $H(z=1)$ on each point of this curve. A
physical solution exists if there is a point in parameter space for
which this value vanishes. A representative case is displayed in
figure~\ref{f:frombdy_ball}.

For the construction of Minkowski solutions, we first determine the
location \mbox{$z=z_{\text{min}}$} at which either $\theta'(z)$ diverges, or
for which $H'(z)$ diverges or $\theta(z)<0$. Closeness to a physical
solution is then measured by the deviation of $\sin\theta(z_{\text{min}})$
from $\pi/2$. Recursing with a finer grid then eventually again leads
to a curve of regular solutions, on which there is at most one point
for which both boundary conditions hold.

\section{Understanding the mass gap and behaviour of ball branes}

Our findings from the previous section are interesting, although
unusual, as they seem to impose a constraint on the \emph{bare} quark
masses, i.e. an ``external'' parameter of the system. In the present
section we would therefore like to see if we can understand the
origin of the gap which the magnetic field has introduced in the mass
spectrum. We will do this by following what is happening with branes
which, in the absence of a magnetic field, have two separated phases.

We would also like to see more explicitly how the ball branes
disappear from the spectrum as the magnetic field is increased, and to
better understand the nature of the maximal value of the magnetic
field $\hat{H}_2$, beyond which no ball brane is present in the
spectrum.  We will do this by following what is happening with the
equatorial ball brane, which is the simplest of all the ball branes
and the last to disappear when the external field reaches its maximal
value.

\subsection{Appearance of the gap: disappearance of critical branes}

In the absence of a magnetic field the two observed phases are
separated by a critical brane, i.e.~a brane which satisfies both
the Minkowski and the ball brane boundary conditions.  So to understand better
the origin of the gap, we will now show \emph{analytically} that these
branes are no longer possible as soon as a magnetic field is introduced.

The boundary conditions which are satisfied by critical branes are given by
\begin{equation}
\chi\left(u=\frac{R}{2} \right) = 1 \, , \quad \quad H\left( u=\frac{R}{2}\right) = 0 \, ,
\end{equation}
since both $S^3$ in $AdS_5$ and $S^3$ in $S^5$ shrink to zero size for
these branes.  Therefore the most general expansion satisfying both of
these conditions is given by
\begin{equation}
\begin{aligned}
\chi(u)&= 1+\chi_1(u-R/2) + \chi_2(u-R/2)^2 + O((u-R/2)^3)\,, \\[1ex]
H(u) &= H_{c1}(u-R/2) + H_{c2}(u-R/2)^2 + O((u-R/2)^3) \, .
\end{aligned}
\end{equation}
By plugging these into the equations of motion, and expanding near the
origin to the second order, one obtains two analytic solutions. For
one of the solutions the variable $\chi$ is manifestly larger than one, and
as such is unacceptable, bearing in mind that $\chi$ is the cosine of
an angular variable. Another solution is given by
\begin{equation}
\begin{aligned}
\chi(u) &= 1-2(u-R/2)^2 + 4(u-R/2)^3 -\frac{52}{15}(u-R/2)^4+ O((u-R/2)^5)\,, \\[1ex]
H(u) &= O((u-R/2)^5) \, .
\end{aligned}
\end{equation}
So we see that the requirement of an analytic solutions of the
equations of motion near the AdS origin implies that branes with a
critical embedding are possible only if there is no magnetic field
turned on (we have verified $H(u)$ vanishes to much higher
order). This fact is suggestive of the formation of a gap between the
two phases, at least in the interior of the $\text{AdS}$ space. Of
course, there is a logical possibility that another type of phase
appears in the gap, which is characterised by boundary conditions or
symmetries which are different from the ones we have assumed so far,
and were therefore missed in the analysis above.  We will shed some
more light on this point when we analyse the fluctuations of the ball
and Minkowski branes in section~\ref{s:fluctuations_and_gap}.

\subsection{Behaviour of the equatorial brane and maximal value 
of magnetic field}

In order to get a better understanding of the disappearance of ball
branes, and of the critical magnetic field $\hat{H}_1$, we will now
focus our analysis on the simplest of ball branes, the equatorial
brane.  As it has the smallest bare mass, this ball brane is the last
one to disappear from the spectrum.

The equatorial brane is the simplest ball brane which exists, and is
it characterised by $m=0$ and $\chi=0$.  In the absence of the
external field, this brane is ``flat'' i.e.~it has a worldvolume
metric $\text{AdS}_5\times S^3$, a submanifold of vanishing extrinsic
curvature.  As the magnetic field is turned on, the shape of this
brane does not change, which is guaranteed by symmetry arguments, and
easily confirmed by equations of motion. However, one still needs to
solve the equations of motion for the magnetic field on the brane
worldvolume.

The action for the magnetic field  is relatively simple and given by 
\begin{multline}
\label{actionequator}
S=-T_{\text{D7}}\frac{R^7}{16}\int\!{\rm d}^8\sigma\, \frac{1}{z^5}
\left(z^2+1\right) \sin\theta\, \cos\theta\, \sin\bar\theta\, \cos
\bar\theta \\[1ex]
\times\sqrt{\left(16 z^4
  H(z)^2+\left(z^2-1\right)^4\right)
  \left(z^4H'(z)^2+\left(z^2-1\right)^2\right)}\,,
\end{multline}
while the equations of motion are
\begin{multline}
H''(z)-\frac{16 z^4 H(z) H'(z)^2}{16 z^4 H(z)^2+\left(z^2-1\right)^4}-\frac{16 \left(z^2-1\right)^2 H(z)}{R^2 \left(16 z^4 H(z)^2+\left(z^2-1\right)^4\right)}
 \\[1ex]
+\frac{\left(16 \left(z^4-4 z^2-1\right) z^4 H(z)^2+\left(z^2-1\right)^4 \left(3 z^4+1\right)\right) H'(z)}{16 \left(z^4-1\right) z^5 H(z)^2+\left(z^2-1\right)^5 \left(z^2+1\right) z} \\[1ex]
+\frac{R^2 z^3 \left(16 z^4 H(z)^2+\left(z^2-1\right)^2 \left(3 z^4+2 z^2+3\right)\right) H'(z)^3}{16 z^4 \left(z^4-1\right) H(z)^2+\left(z^2+1\right) \left(z^2-1\right)^5}=0\,.
\end{multline}
These equations still have to be solved numerically. However, we are
now dealing only with two free parameters $(H_{\text{ext}},M)$. So we can
solve them by shooting either from infinity or from the origin,
and in that way cross check our findings.
 
Observe that the action for equatorial brane (\ref{actionequator}) is
always real for all values of the magnetic field. In other words, the
reality condition on the action does not imply the existence of a
maximal value of the external magnetic field, but this will have to
come out from solving equations.

\begin{figure}[th]
\label{solnsequatorial}
\begin{center}
\includegraphics[width = 0.48\textwidth]{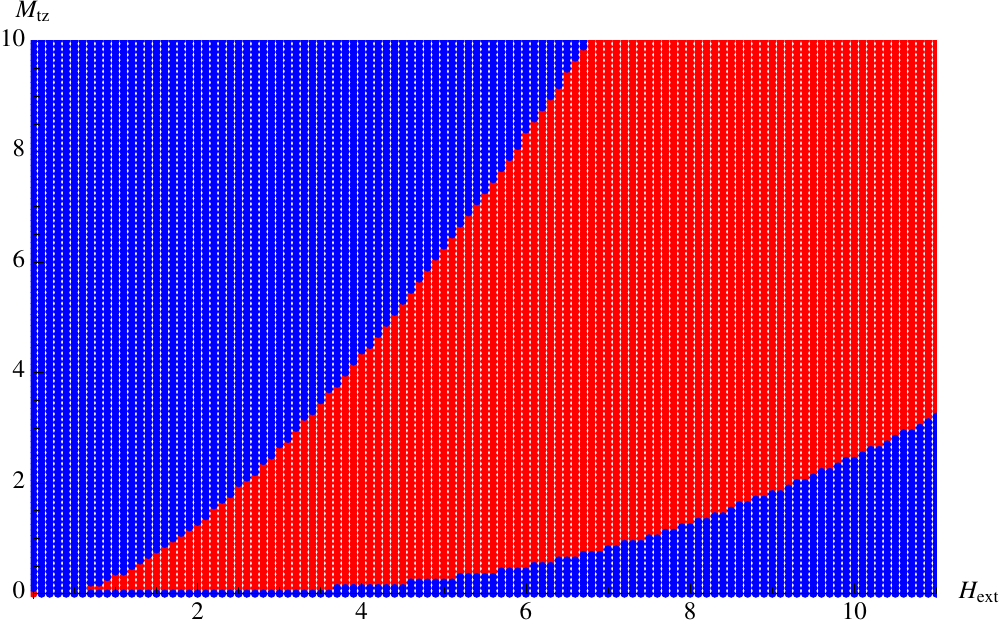}
\includegraphics[width = 0.48\textwidth]{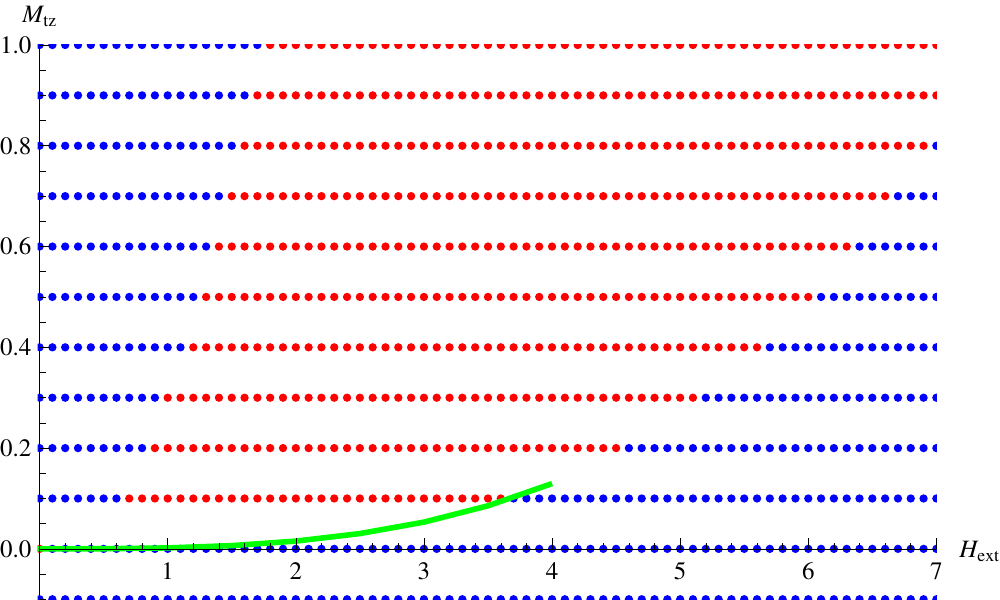}
\end{center}
\caption{The plot shows the $(H_{\text{ext}},M_{tz})$ parameter space.
  In the red (middle) region, the value of the magnetic field at the
  origin is finite, but non-vanishing. In the blue (surrounding)
  regions the derivative of the magnetic field blows up as one
  approaches the origin of the AdS space. The physical solutions are
  located on the borderline, between the two regions, as shown in the
  plot on the right hand side (green curve).}
\end{figure}

For generic values of the parameters $(H_{\text{ext}},M_{tz})$ at
infinity, the magnetic field of the solution at the origin of the AdS
space has a non-vanishing value, or its derivative blow up. However,
for the points on the green line of figure (\ref{solnsequatorial}) the
magnetic field satisfies the required boundary condition at the origin
and it is regular everywhere. We see that this green line of physical
solutions is \emph{part} of the borderline between the red region
(where the magnetic field has non-vanishing value at the origin) and
the blue region (where the derivative of the magnetic field blows up in
the interior of AdS space).

By looking at right hand side plot of figure (\ref{solnsequatorial})
we see that the green line of physical solutions stops for some
``critical'' value of the external field $H_{\text{ext}} = \hat{H}_1
\approx 3.98$.  This signals that the equatorial Minkowski brane does
not exist for a value of magnetic field which is larger than the value
$\hat{H}_1$.

\section{Fluctuation analysis and investigation of the gap}
\label{s:fluctuations_and_gap}

The analysis in the previous section suggests that the magnetic field
introduces a gap in the spectrum and leads to the ball branes
disappearing from the spectrum. However, there is a possibility that
the branes which were analysed in the previous sections were not
general enough, and that the appearance of the gap is a consequence of
the fact that our ansatz was not general enough.

Therefore, in order to gain some insight into this question, it would
as a first step be useful to understand the stability of the
constructed branes under small fluctuations. In particular, we would
like to understand the stability of branes near the gap. In the
present section we therefore initiate an analysis of the stability of
branes near the gap, and discover an instability of the branes under
the fluctuations of the transverse scalar mode. We then try to use
this knowledge to shed some light on the potential branes which could
fill out the gap.

\subsection{Stability analysis of the branes}

The stability analysis of a generic ball or Minkowski brane proves to
be a very challenging numerical problem. The reason is that, to start
with, all branes are known only numerically (except for the equatorial
brane). In addition, for a generic brane, the transverse scalars and
vector modes mix.  The fluctuation analysis in \cite{Filev:2012ch} was
done with a background magnetic field which does not satisfy the
equations of motion (as already discussed in the
section~\ref{s:EOM}). Moreover, instead of first deriving the
equations of motion for fluctuations and then plugging in the ansatz
for the fluctuations, the authors of~\cite{Filev:2012ch} inserted the
fluctuation ansatz directly into the action, thus failing to check if
the ansatz for the fluctuations is consistent with full equations of
motion. Following the full path of first deriving equations of motion
for the arbitrary brane near the gap is very involved, both
analytically and numerically, as unfortunately the analysis does not
seem to simplify near the gap. Fortunately, the analysis for the
equatorial brane, which reaches the gap for $H_{\text{ext}}\sim 3.98$,
is much simpler, and can be done in full generality.  For this brane,
the shape is simple and known explicitly.  So in what follows we will
focus on the equatorial brane, for a generic value of the magnetic
field.

When studying the fluctuations around the configurations constructed in
the previous sections, we will need the Wess-Zumino term in addition to
the DBI action,
\begin{multline}
S_{\text{WZ}} = \frac{1}{2} T_{\text{D7}}(2\pi\alpha')^2\int P[C_4]\wedge F\wedge F+
\frac{1}{2} T_{\text{D7}}\int P[C_4]\wedge P[B]\wedge P[B]\\[1ex]
+T_{\text{D7}} 2\pi\alpha'\int P[C_4]\wedge P[B]\wedge F,
\end{multline}
where $C_4$ is a 4-form gauge field with self-dual
field strength $F_5$ given by
\begin{equation}
F_{5} = \frac{4}{R} \left(\text{Vol}(\text{AdS}_5)+\text{Vol}(S^5) \right) \, .
\end{equation}
The equations of motion following from the full  action are
given by
\begin{equation}
\label{EOMgfull}
\partial_a(\sqrt{-{\cal E}}({\cal E}^{ab}-{\cal E}^{ba})) = -\frac{1}{5!} F_{a_1a_2a_3a_4a_5}{\cal F}_{a_6a_7}\tilde{\epsilon}^{a_1\cdots a_7b}
\end{equation}
and
\begin{multline}
\label{EOMefull}
-\frac{1}{4\cdot 4!}\tilde{\epsilon}^{a_1\cdots a_8}F_{\rho a_1a_2a_3a_4}{\cal F}_{a_5a_6}{\cal F}_{a_7a_8}=\partial_b\left(\sqrt{-{\cal E}}({\cal E}^{ba}+{\cal E}^{ab})\right) G_{\nu\rho}\partial_a x^\nu  \\[1ex]
+2 \sqrt{-{\cal E}}{\cal E}^{ba} \left(G_{\nu\rho}\partial_b\partial_a x^\nu+\frac{1}{2} \left( \partial_\mu G_{\rho\nu} - \partial_\rho G_{\mu\nu} + \partial_\nu G_{\mu\rho} \partial_a x^\mu\partial_b x^\nu \right) \right)\,,
\end{multline}
where ${\cal F}=P[B]+2\pi\alpha'F$.  Greek indices are spacetime
indices while Roman indices are worldvolume indices.  For the
spacetime fields having worldvolume indices, it is understood that the
fields are pulled back to the worldvolume.

Recall that the equatorial brane is specified by
\begin{equation}
\chi = 0 \, , \quad \kappa = 0\,,
\end{equation}
together with
\begin{multline}
B = \frac{1}{2} H'(u)\, R^2(\sin^2\bar{\theta}\, {\rm d}u {\rm d}\bar{\phi}+\cos^2
\bar{\theta}\, {\rm d}u {\rm d}\bar{\psi})\\[1ex]
+ H(u)\, R^2 \sin \bar{\theta}\cos\bar{\theta}\,({\rm d}\bar{\theta} {\rm d}\bar{\phi}- {\rm d}\bar{\theta} {\rm d}\bar{\psi}) \, .
\end{multline}
The fluctuations around the equatorial brane consist of scalars which
are transverse to the brane worldvolume (these are not charged under
the $SO(4)$ isometry group of the $S^3 \in S^5$ which is wrapped by
the D7-brane), vector fluctuations in the direction of $S^3 \in S^5$
(which are dual to scalars from the field theory point, we call these
``charged scalars'') and vector fluctuations of the gauge field in the
non-compact directions of the probe brane.

In order to study charged scalar fluctuations, we make the following ansatz,
\begin{equation}
\delta A_i =  e^{-i\omega t}a_{\omega,\bar{l},l,s}(u)\,Y^{\bar{l}}(\bar{\Omega}_3)\,Y_i^{l,s}(\Omega_3)\,,
\end{equation}
where $Y^{\bar{l}}(\bar{\Omega}_3)$ are spherical harmonics,
$Y_i^{l,s}$ are vector spherical harmonics, with $l,\bar{l}$ being
integer and $s = \pm 1$.  From the previous analyses in
\cite{Erdmenger:2010zm,Chunlen:2012zy}, it is known that the mode with
$l=1,s=-1, \bar{l}=0$ has the lowest energy in the absence of a
magnetic field. Since we do not expect modes to cross when the
magnetic field is turned on, if there is an instability it will first
show up in this mode of fluctuations. 
Therefore, in what follows we will focus on this fluctuation
mode. The ansatz then becomes explicitly,
\begin{equation}
\delta A = a(t,u)(\sin^2\theta {\rm d}\phi+\cos^2\theta {\rm d}\psi)
\,, \quad a(t,u)= e^{-i \omega t} a(u) \, .
\end{equation}
The equation of motion for the charged scalar fluctuation $a(u,t)$ is 
\begin{multline}
\label{e:charged_scalar_fluct}
 a''(u)+\frac{\partial_u\left(\sqrt{-e}e^{uu}\right)}{\sqrt{-e}e^{uu}}a'(u)+\frac{1}{e^{uu}}\left(\frac{4}{R^2}-e^{tt}\omega^2\right)a(u)  \\[1ex]
 -\frac{2}{R^2}\left(4g_{uu}+R^4H'(u)^2g^{33}-4\sqrt{\frac{g_{uu}e^{33}}{e^{uu}g^{33}}}\right)a(u)=0\,.
\end{multline}
This equation is very similar to the equations
obtained in~\cite{Chunlen:2012zy} for the analysis of the ground state
of $N=2$ theory on $S^3$ in the presence of an isospin chemical
potential. The equation is solved by putting it into Schr\"odinger
form. For all the technical details, we refer the reader
to~\cite{Chunlen:2012zy}.

\begin{figure}[th]
\begin{center}
\vspace{2ex}
\includegraphics[width = 0.52\textwidth]{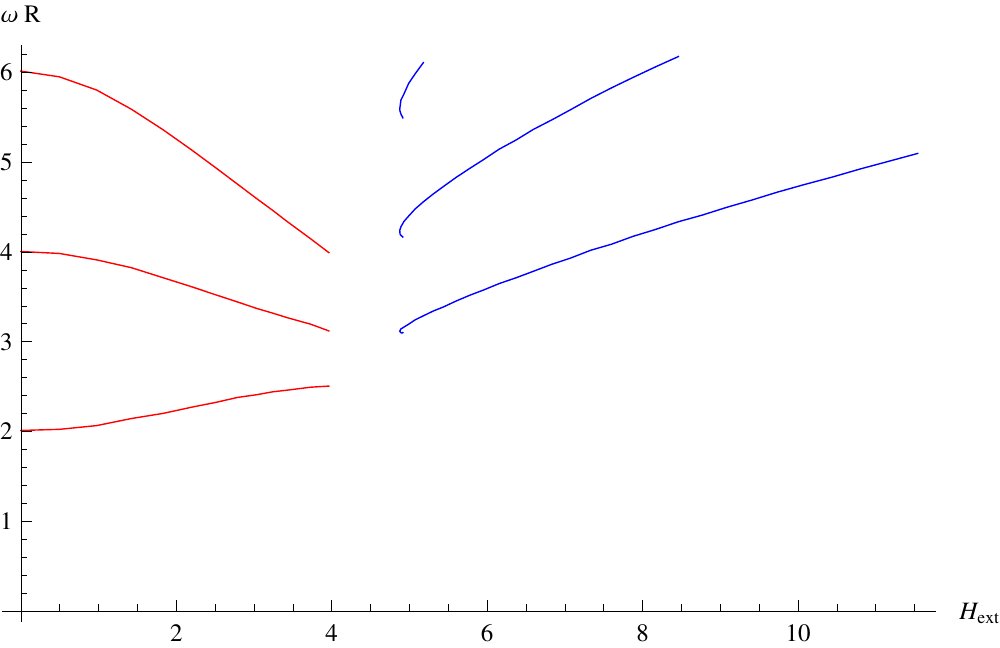}
\vspace{-2ex}
\end{center}
\caption{The first three normal modes of the charged scalar
  fluctuation~\eqref{e:charged_scalar_fluct} for fixed $m=0$, as a function
  of the external magnetic field.  Red shows the plot for ball
  embeddings while blue shows the plot for Minkowski
  embeddings. \label{fluctsvectorsphere}}
\end{figure}

\begin{figure}[th]
\vspace{2ex}
\begin{center}
\includegraphics[width = 0.42\textwidth]{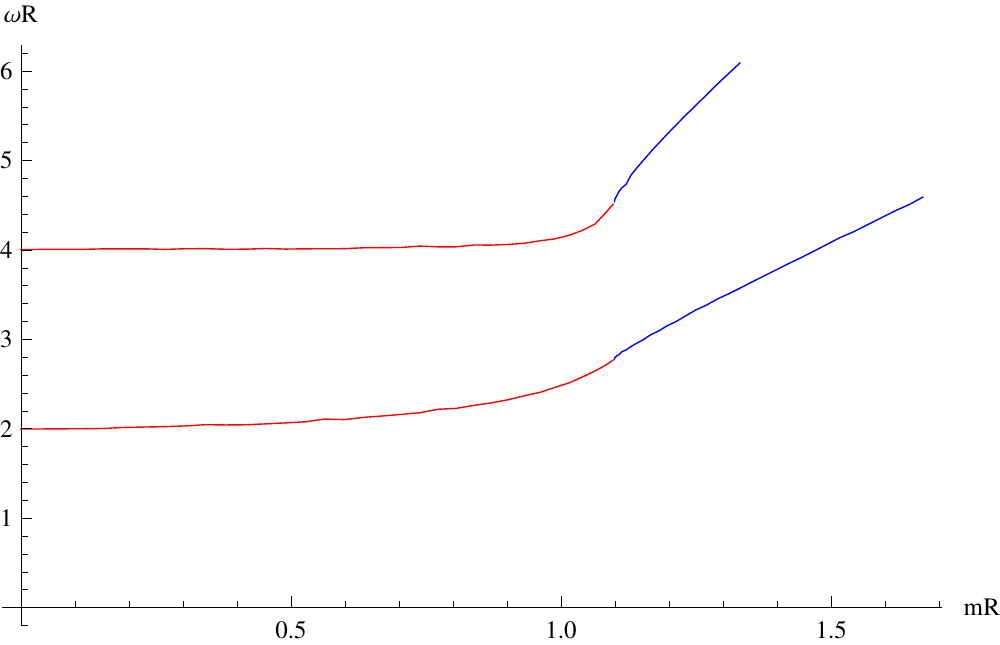}\qquad
\includegraphics[width = 0.42\textwidth]{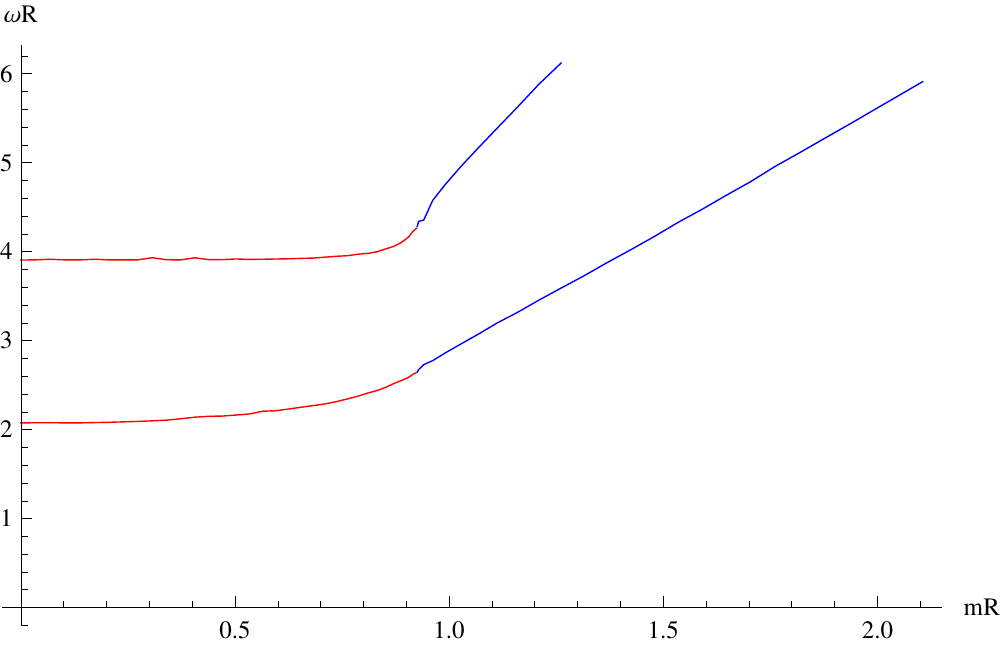}\\[1ex]
\includegraphics[width = 0.42\textwidth]{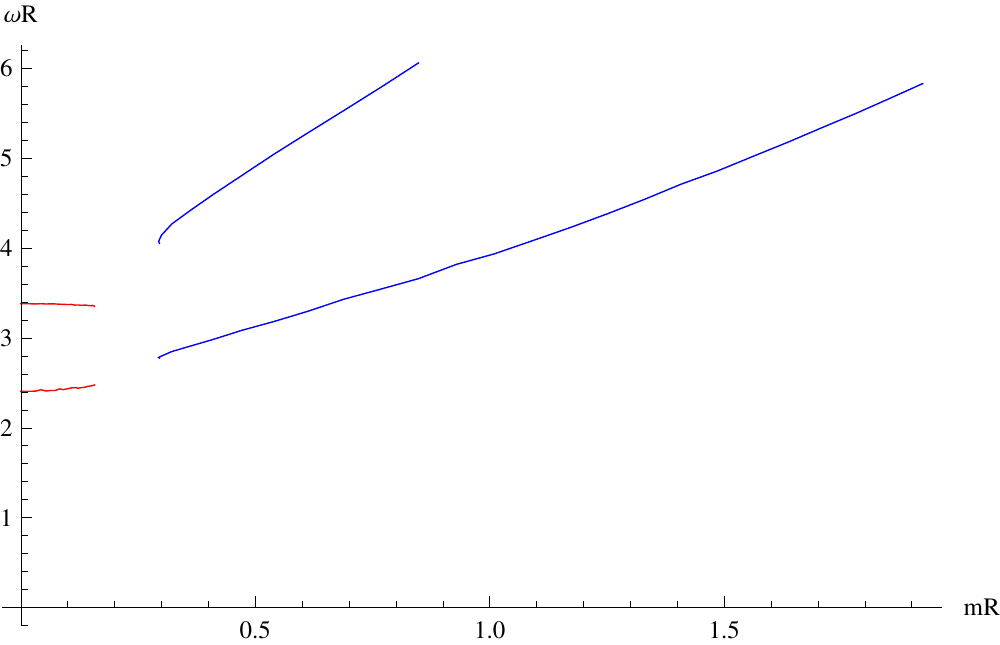}\qquad
\includegraphics[width = 0.42\textwidth]{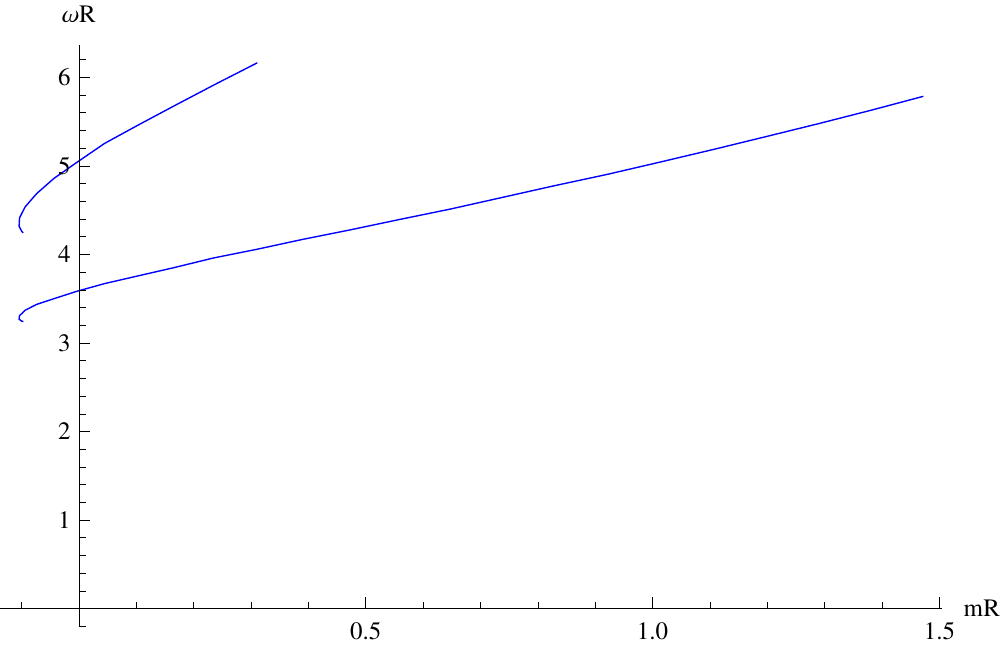}
\end{center}
\caption{The first two normal modes of the charged scalar
  fluctuation~\eqref{e:charged_scalar_fluct} for fixed external magnetic field
  but varying quark mass.  From left to right and top to bottom,
  $H_{\text{ext}} = 0,1,3,6,$ respectively.  Red shows the plot for
  ball embeddings while blue shows the plot for Minkowski
  embeddings. \label{fluctuationspherebig} }
\end{figure}

Solving the equations of motion for the fluctuations determines the
frequencies of the fluctuations as functions of the external magnetic
field. We impose the same boundary conditions for fluctuations as for
the brane around which the fluctuations are excited. The solutions of
the equations are labelled by an `$n$', a new integer quantum number.
Figure \ref{fluctsvectorsphere} shows the frequencies of the first
three modes $n=1,2,3$. We see that despite the fact that the frequencies
for the equatorial ball brane decrease as the external field
increases, they remain positive even when all ball branes disappear. A
similar analysis can be performed for non-vanishing quark masses, and
a similar behaviour is observed for all other branes as shown in
figure~\ref{fluctuationspherebig}. We thus see no signs of instability
under the fluctuations in the direction of the maximal $S^3 \in S^5$.

\begin{figure}[th]\vspace{2ex}
\begin{center}
\includegraphics[width = 0.48\textwidth]{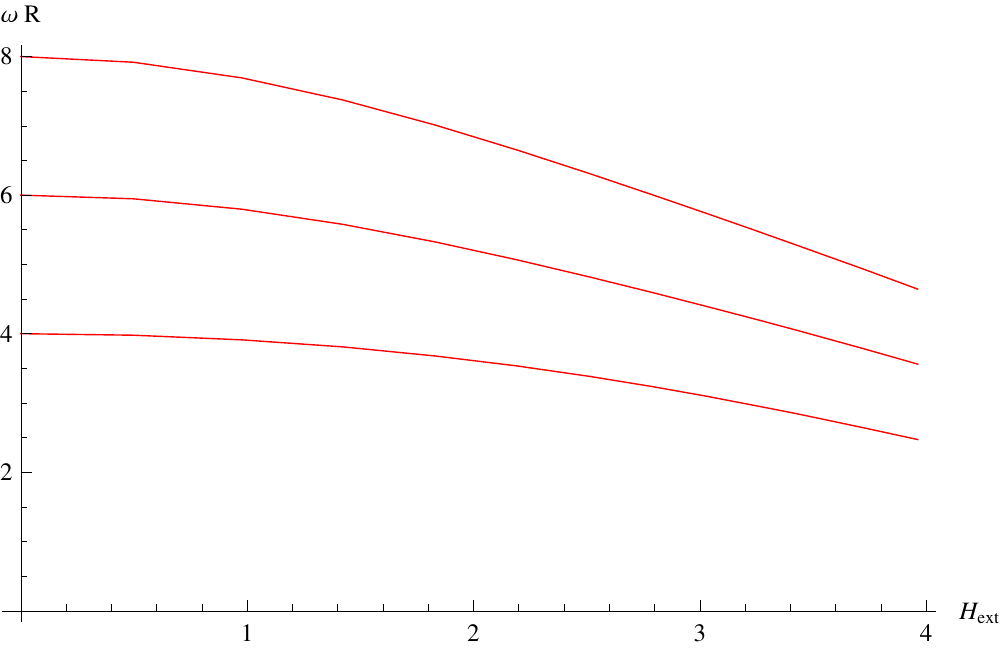}
\end{center}
\vspace{-2ex}
\caption{First three normal modes of the vector
  fluctuation~\eqref{e:vector_fluct} for the equatorial embedding $m =
  0$ as a function of the external field.
\label{fluctuationsvect}
}
\end{figure}

Next we turn to the analysis of vector fluctuations in the direction
of the AdS part of the brane worldvolume. This analysis is similar to
the previous analysis of the charged scalar, except that because the
magnetic field is non-vanishing in the same directions as those in
which we fluctuate (i.e.~the $\bar{S}^3 \in AdS_5$) the energy levels
will in general split. Therefore, the magnetic quantum numbers should
be also taken into account, so that one is in general dealing with
$Y^{l, m_1,m_2,s}$. The general ansatz we make is 
\begin{equation}
\delta A_i =  e^{-i\omega t}a_{\omega,\bar{l},\bar{m}_1,\bar{m}_2, \bar{s}}(u)\bar{Y}_i^{\bar{l},\bar{m}_1, \bar{m}_2,\bar{s}}(\bar{\Omega}_3) \, .
\end{equation}
While we have performed the analysis for general $m_1,m_2$ it turns out
that $(\bar{l},\bar{m}_1,\bar{m}_2,\bar{s})=(1,0,0,-1)$ is the lowest mode.
The ansatz for the fluctuation given previously simplifies for the lowest
lying mode to
\begin{equation}
\delta\bar{A} = \bar{a}(t,u)(\sin^2\bar\theta {\rm
  d}\bar\phi+\cos^2\bar\theta {\rm d}\bar\psi)\,,\quad
\bar{a}(u,t) = e^{-i \omega t} \bar{a}(u)\,.
\end{equation}
The equation of motion for the vector fluctuation $\bar{a}(u,t)$ is 
\begin{multline}
\label{e:vector_fluct}
\bar{a}''(u) + \frac{\partial_u(\sqrt{-e}(e^{uu})^2g_{uu}g^{33})}{\sqrt{-e}(e^{uu})^2g_{uu}g^{33})}\bar{a}'(u) - \frac{\omega^2 e^{tt}+4e^{33}}{e^{uu}}\bar{a}(u) \\[1ex]
+\left(2H'(u)\frac{\partial_u g^{33}}{g_{uu}(g^{33})^2}+8H(u)\right)H(u)R^4(e^{33})^2\frac{g^{33}}{e^{uu}}\bar{a}(u)= 0 \, .
\end{multline}
After an analysis similar to the one for the charged scalar, we get
the frequencies as a function of the external field. These are presented in
figure~\ref{fluctuationsvect}. We again observe that all modes stay
positive definite, even if the magnetic field exceeds it critical value.

Let us finally turn to the scalar fluctuations. We focus on the equatorial
embedding of the D7-brane. It is convenient to introduce coordinates on $S^5$
as
\begin{equation}
w_1=R\sin\theta\cos\phi \, \quad w_2=R\sin\theta\sin\phi ,
\end{equation}
so that the metric  on $S^5$ becomes 
\begin{equation}
\label{S5metrics}
{\rm d}s_5^2 = \left(1-\frac{w_1^2+w_2^2}{R^2} \right) {\rm
  d}\Omega_3^2 + \frac{1}{1 - \frac{w_1^2 +w_2^2}{R^2}} \bigg( {\rm d}w_1^2
+ {\rm d}w_2^2 - \left(\frac{w_2}{R}{\rm d} w_1 - \frac{w_1}{R}{\rm d} w_2\right)^2
\bigg) \,.
\end{equation}
The embedding of the equatorial brane is given by
\begin{equation}
w_1 = w_2 = 0 \, .
\end{equation}
Hence, we make the following ansatz for the fluctuations,
\begin{equation}
\delta w_\mu =  2 \pi \alpha' \Psi_\mu(u) \,, \quad \mu = 1,2 \, .
\end{equation}
The fluctuation equations are symmetric with respect to $w_1$ and
$w_2$, and moreover, it is consistent to set either of the two scalars
to zero value.  So we choose the following ansatz for the fluctuations
\begin{equation}
\Psi_1 = \Psi(u) e^{-i \omega t} \, , \quad  \Psi_2 = 0\,.
\end{equation}
Substituting this ansatz into the equations of motion, and linearising
in the fields, one obtains the equation governing the dynamics of $\Psi$,
\begin{equation}
\label{e:scalar_fluct}
\partial_u^2 \Psi(u)+\frac{\partial_u \left( \sqrt{-e}  e^{uu} \right) } {\sqrt{-e} e^{uu}} \partial_u\Psi(u)+\frac{1}{ e^{uu}} \left(\frac{3}{R^2}-  e^{tt}\omega^2\right) \Psi(u)=0 \,  .
\end{equation}
where 
\begin{align}
\sqrt{-e} &= \frac{(R^2 + 4 u^2) \sqrt{\bigg(256 R^4 u^4 H(u)^2 + (R^2 - 4 u^2)^4 \bigg) \bigg( 4 R^2 u^4 H'(u)^2  + (R^2 - 4 u^2)^2 \bigg) }}{u^5} \,,\nonumber\\[1ex]
e^{uu} &=  \frac{(R^2 u - 4 u^3)^2} {4 R^4 u^4  H'(u)^2 + (R^3 - 4 R u^2)^2}\,, \\[1ex]
e^{tt} &= - \frac{16 R^2 u^2}{(R^2 + 4 u^2)^2} \, .\nonumber
\end{align}

\begin{figure}[th]
\begin{center}
\includegraphics[width = 0.48\textwidth]{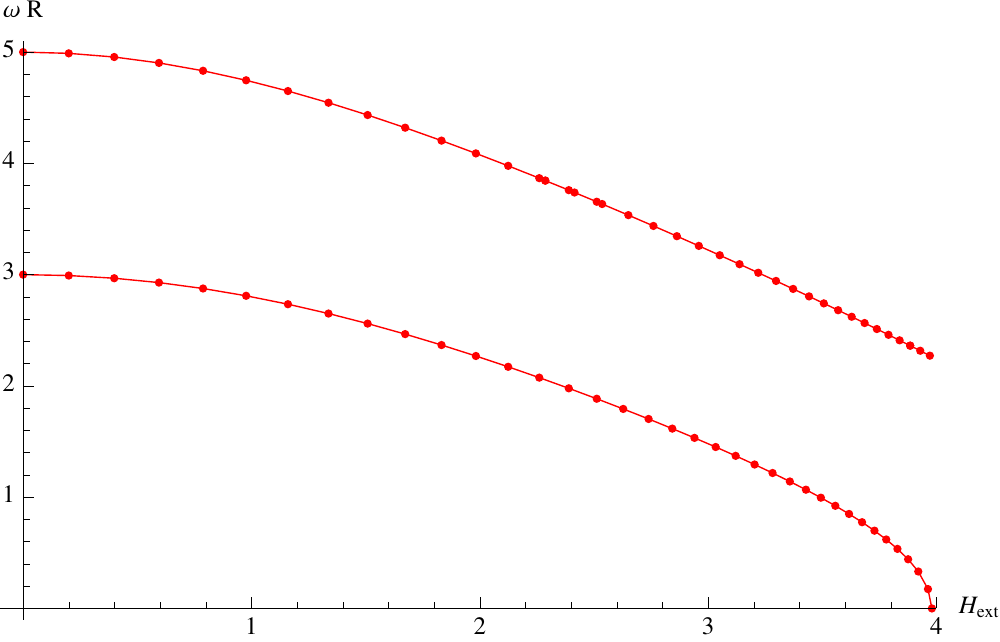}
\includegraphics[width = 0.48\textwidth]{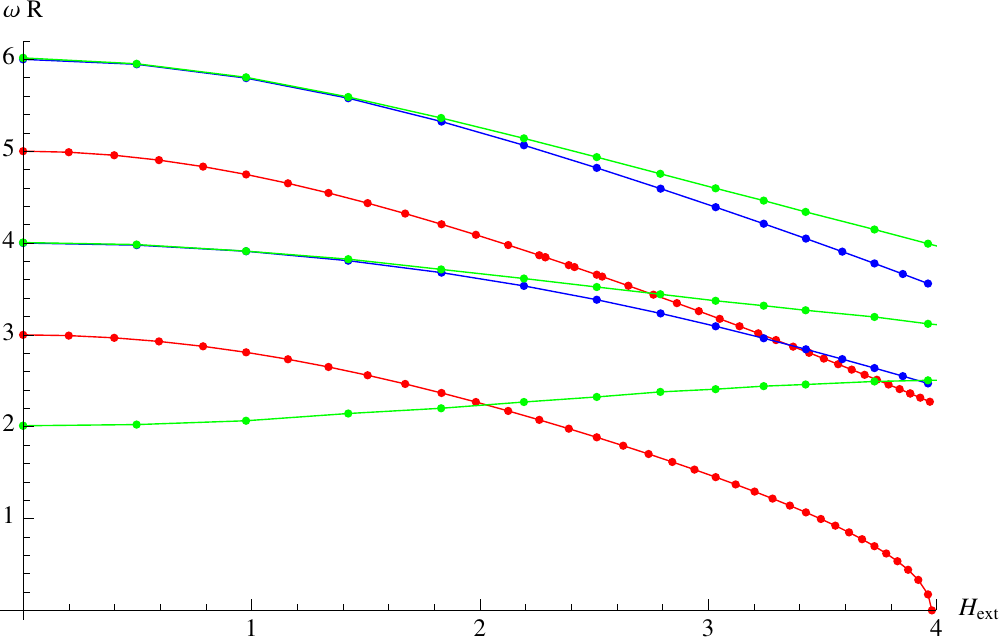}
\end{center}
\caption{The panel on the left shows the first two normal modes of the
  scalar fluctuation~\eqref{e:scalar_fluct} for the equatorial
  embedding $m = 0$ but varying external magnetic field.  The panel on
  the right shows these modes together with the charged scalar modes
  of figure~\ref{fluctsvectorsphere} (green) and the vector modes of
  figure~\ref{fluctuationsvect} (blue) plotted together. We see that
  the magnetic field makes the scalar fluctuation unstable.  }
\label{scalarfluctuations}
\end{figure}

By solving this equation we get the frequencies as functions of the
external magnetic field and $n$, a new quantum number. These functions
are shown in figure~\ref{scalarfluctuations} for the first two
lightest modes. We see that for large enough value of the magnetic
field, $H\sim 3.98$, the lightest scalar mode becomes unstable. The
value for which this happens is exactly the same as the value for
which the equatorial ball brane disappears from the spectrum. So this
perturbative analysis is consistent with our previous analysis, as it
suggests that the real ground state for $H>3.98$ is no longer a ball
brane.

\subsection{Investigating the gap}

An unusual feature of the gap which was discovered in the first part
of the paper is that it introduces restrictions on the bare parameters
of the theory. The gap in the spectrum forbids certain values of the
bare quark masses, depending on the value of the external field. While
this is in principle possible, we would like to eliminate the
possibility that the presence of the gap is due to missed states with
a different symmetry, which we have not allowed for in our analysis so
far. While we do not yet have a conclusive answer to this question, we
will present here our attempts to understand if there are additional
states present in the ``gap'' part of the spectrum.

One observation which we make is that if one relaxes the condition
that the magnetic field has to vanish at the origin of the AdS space,
then the gap can be filled completely with ball branes. Of course,
these branes are unphysical as the magnetic field is non-vanishing in
the direction of the $S^3$ which shrinks to zero size at the origin of
$\text{AdS}_5$. However, perhaps there is a way in which this kind of
singular behaviour could be resolved by ``blowing up'' the shrinking
$S^3$. We will leave this line of investigation for later and
focus on an alternative route here.

The analysis of the fluctuations from the previous section suggests
that branes near the gap are unstable under transverse scalar
fluctuations.  This suggests that near the branes which are at the
edge of the gap, there could be a new, nearby ground state, which has
less symmetry than either ball or Minkowski branes. In particular,
since the unstable fluctuation breaks the $\bar{SO}(4)$ symmetry
preserved by ball and Minkowski branes, we expect that the new ground
state should also break this symmetry.  Making a general ansatz for
the transverse scalar $\chi$, where it depends on all coordinates of
the $\bar{S}^3 \in \text{AdS}_5$, gives a very complicated set of equations
which we do not know how to analyse in full generality.  Therefore, we
will make the simplifying assumption that the real ground state is
infinitesimally close to the equatorial brane, and use this to attempt
to construct the new ground state perturbatively. A similar analysis
was made (for a different model) in~\cite{Bu:2012mq}, who constructed
the Abrikosov lattice in a holographic framework. The perturbative
parameter is the distance from the critical magnetic field for which
the gap appears,
\begin{equation}
\Delta = \frac{H_{\text{ext}} - \hat{H}_1}{\hat{H}_1}
\end{equation}
where $\hat{H}_1=3.98$ for the equatorial brane.

We make the following ansatz
\begin{equation}
\label{expansionextra}
\begin{gathered}
\delta \chi(u,\bar{\theta},\bar{\phi},\bar{\psi}) = \Delta \chi_1(u,\bar{\theta},\bar{\phi},\bar{\psi}) + \Delta^3\chi_3(u,\bar{\theta},\bar{\phi},\bar{\psi}) + O(\Delta^5) \\[1ex]
\delta \Lambda = \Delta^2 \Lambda_2 + \Delta^4 \Lambda_4 + O(\Delta^6)
\, ,  \quad   \delta B = {\rm d} \delta\Lambda
\end{gathered}
\end{equation}
Here, alternating of the powers of the expansion in $\delta \chi$ and
$\delta \Lambda$ is motivated by~\cite{Bu:2012mq} and in turn by the
original paper~\cite{Abrikosov:1956sx}. Plugging this ansatz into the
equations of motion we can solve the equations order by order. We omit
the details of this lengthy analysis here, but the upshot is that at
third order in perturbation theory the equation for $\chi_3$ does not
seem to have a physical solution.

While this attempt does not seem to lead to a new ground state, we
still believe that making some modifications of this approach will
work.  In particular, relaxing the alternating powers in the
expansion~\eqref{expansionextra} may help to resolve the issues which
we observed. It may also be necessary to turn on a non-abelian
component of the worldvolume gauge field, as was done in
\cite{Bu:2012mq}.  We intend to address these issues in future work.

\vfill\eject

\providecommand{\href}[2]{#2}\begingroup\raggedright\endgroup

\end{document}